\documentclass[twocolumn,pra,showpacs,superscriptaddress]{revtex4-1}
\usepackage{graphicx}
\usepackage{amssymb,amsmath}
\usepackage{bm}
\usepackage{dcolumn}
\usepackage{subfigure}
\usepackage{float}
\usepackage{url}
\usepackage{color}
\usepackage{txfonts}
\usepackage[driverfallback=dvipdfm,CJKbookmarks,pdfpagelabels]{hyperref}
\hypersetup{backref,pdfpagemode=FullScreen,colorlinks=true,breaklinks,urlcolor=blue,linkcolor=blue,citecolor=blue}

\begin{document}
\title{Generating topological optical flux lattices for ultracold atoms\\
by modulated Raman and radio-frequency couplings}
\author{Jinlong Yu}
\altaffiliation{Current address: Institute for Advanced Study, Tsinghua University, Beijing 100084, China}
\affiliation{State Key Laboratory of Low Dimensional Quantum Physics, Department of Physics, Tsinghua University, Beijing 100084, China}
\author{Zhi-Fang Xu}
\email{zfxu83@gmail.com}
\affiliation{MOE Key Laboratory of Fundamental Physical Quantities Measurements, School of Physics, Huazhong University of Science and Technology, Wuhan 430074, China}
\author{Li You}
\email{lyou@mail.tsinghua.edu.cn}
\affiliation{State Key Laboratory of Low Dimensional Quantum Physics, Department of Physics, Tsinghua University, Beijing 100084, China}
\affiliation{Collaborative Innovation Center of Quantum Matter, Beijing 100084, China}

\begin{abstract}
 We propose a scheme to dynamically generate optical flux lattices with nontrivial band topology using amplitude-modulated Raman lasers and radio-frequency (rf) magnetic fields.
 By tuning the strength of Raman and rf fields, three distinct phases are realized at unit filling for a unit cell. Respectively, these three phases  correspond to normal insulator, topological Chern insulator, and semimetal. Nearly nondispersive bands are found to appear in the topological phase,
 which promises opportunities for investigating strongly correlated quantum states  within a simple cold-atom setup.
 The validity of our proposal is confirmed by comparing the Floquet quasienergies from the evolution operator with the spectrum of the effective Hamiltonian.
\end{abstract}


\maketitle

\section{Introduction}

Orbital magnetism or spin-orbit coupling (SOC) is essential for exotic quantum states such as topological insulators~\cite{Hasan2010, Qi2011}. For charge neutral ultracold atoms, orbital magnetic field can emerge either in the noninertial frame by rotating atomic quantum gases~\cite{Madison2000, Abo-Shaeer2001}, or through engineered  adiabatic motion in Raman coupled atomic states with spatially varying two-photon detuning~\cite{Lin2009}.
By using Raman lasers to couple atomic states, where atomic pseudo-spin flip is always accompanied by a momentum change, the one-dimensional (1D) SOC with equal Rashba and Dresselhaus strength can be realized~\cite{Lin2011}. Such a 1D SOC was  demonstrated first by the NIST group for bosonic atoms~\cite{Lin2011} and subsequently by the SXU group~\cite{Wang-Zhang2012} and MIT group~\cite{Cheuk-Zwierlein2012} for fermionic systems. We will call it the 1D Raman-induced SOC, following Ref.~\cite{Zhai2015}.
Two-dimensional (2D) SOC with elaborately designed Raman lasers is also realized recently~\cite{Huang-Zhang2016, USTC_SOC2016}. Apart from using Raman lasers, the generation of 1D SOC by modulated gradient magnetic fields is also demonstrated experimentally both in free space~\cite{Luo2015} and in optical lattices~\cite{Struck_Sengstock2014, Jotzu_Esslinger2015}. Moreover, it is proposed that 2D SOC can be generated in a pure magnetic way by using modulated gradient magnetic fields~\cite{Xu2013, Anderson2013}.

The 1D Raman-induced SOC~\cite{Lin2011, Wang-Zhang2012, Cheuk-Zwierlein2012} embodies an intrinsic spatial structure, as the SOC term explicitly shows up in a frame with space-dependent pseudo-spin rotation~\cite{Lin2011, Zhai2015}.  This fact becomes clear when a radio-frequency (rf) magnetic field, in addition to the Raman lasers, is applied to couple the pseudo-spin  states, which results in  a 1D Zeeman lattice as shown first for bosonic $^{87}$Rb atoms~\cite{Jimenez2012} and subsequently for fermionic $^6$Li atoms~\cite{Cheuk-Zwierlein2012}. In this Zeeman lattice, as the spin states are dressed by Raman and rf fields in a space-periodic manner, an effective spatially periodic 
(orbital) magnetic field emerges for atomic  center-of-mass (c.m.) motion in the adiabatic limit, leading to complex Peierls phase factors in the hopping constants in the tight-binding regime. Thus a scalar lattice potential and a gauge potential are generated simultaneously.
For a 1D Zeeman lattice, the effect of the complex Peierls factors is a simple shift of energy spectrum in the quasimomentum space, which can be gauged away by a redefinition of the creation and annihilation operators.
Nevertheless, generalizations to the 2D cases, such as optical flux lattices~\cite{Cooper2011} and topological magnetic lattices~\cite{Yu2016}, feature nontrivial effects.

The optical flux lattice, which is introduced by Cooper in Ref.~\cite{Cooper2011}, can be viewed as a special type of a 2D Zeeman lattice. The three components of the effective Zeeman field take a nontrivial winding pattern in real space, giving rise to nonzero net flux, hence possibly extremely large flux density for the c.m. degree of freedom in the adiabatic limit. The motion of charged particles in a lattice in the presence of a large external magnetic field~\cite{Hofstadter1976} thus can be simulated by ultracold atoms in an optical flux lattice, and the appearance of quantized transport~\cite{Dauphin2013, Aidelsburger2015} as well as fractional quantum Hall states~\cite{Cooper2013} is expected. It is worth mentioning that alternative approaches to realizing flux lattices with laser assisted tunneling technique~\cite{Jaksch2003} or SOC in a synthetic dimension~\cite{Celi-Lewenstein2014} have been implemented in several recent experiments~\cite{Aidelsburger2013, Miyake2013, Aidelsburger2015, Kennedy_Ketterle2015, Mancini_Fallani2015, Stuhl_Spielman2015}; the realization of strongly correlated fractional quantum Hall states in these systems, however, remains an ongoing task.

The scheme of dynamically generating 2D SOC by gradient magnetic field pulses in Ref.~\cite{Xu2013} can also be generalized to synthesize 2D square magnetic lattices~\footnote{The terminology \emph{magnetic lattice} is basically equivalent to \emph{Zeeman lattice}. Some may use the former to emphasize that such a lattice is generated magnetically.} as shown in~\cite{Luo2014}. With the particular design of the pulse sequence, it is found later that magnetic lattices with nontrivial band topology can be realized~\cite{Yu2016}. Such a topological magnetic lattice shares many similarities with the optical flux lattice, albeit the creation mechanism is different:
The former is realized by modulated magnetic fields~\cite{Yu2016}, while the later relies on
bichromatic laser fields~\cite{Cooper2011b, Juzeliunas2012}. With the observation that the effective Hamiltonian generated by a pair of opposite gradient (or uniform) magnetic fields~\cite{Luo2014, Yu2016} takes a form similar to the one generated by Raman (or rf) fields~\cite{Lin2011,Jimenez2012}, we find that the scheme of generating topological magnetic lattices proposed in Ref.~\cite{Yu2016} can be recomposed to synthesize topological optical flux lattices by modulated Raman and rf fields.

To be more specific, we propose in this paper a scheme to generate a topological optical flux lattice based on the experimental technique for creating a 1D Zeeman lattice~\cite{Jimenez2012, Cheuk-Zwierlein2012}. We show that, by creating three 1D Zeeman lattices with the directions of the corresponding spatial periodicities taking mutual angles of $120^\circ$
in three subsequent time intervals, and supplemented by additional time-periodic $2\pi/3$ rf pulses, an optical flux lattice with nontrivial band topology can be realized dynamically. The topological property of the resulting  lowest energy band of the time-independent effective Hamiltonian depends on the Raman and rf coupling strength. The ground state phase diagram at unit filling as a function of these two parameters is explored. Three different phases are identified through investigating the lowest energy gaps and their associated Chern numbers~\cite{TKNN1982}.
The three phases we find  correspond, respectively, to  topological Chern insulator, normal insulator, and semimetal.

The paper is organized as follows. In Sec.~\ref{SecII}, we review the experimental protocol for generating a 1D Zeeman lattice by Raman and rf fields in Refs.~\cite{Jimenez2012, Cheuk-Zwierlein2012}.
In Sec.~\ref{SecIII}, we describe our proposal for generating topological optical flux lattices.  In Sec.~\ref{SecIV}, we identify the ground state phase diagram and explore topological properties of the different phases. The validity of our proposal is also discussed. We conclude in Sec.~\ref{SecV}.

\section{The protocol for generating a 1d Zeeman lattice}\label{SecII}

\begin{figure}
  \includegraphics[width=\columnwidth]{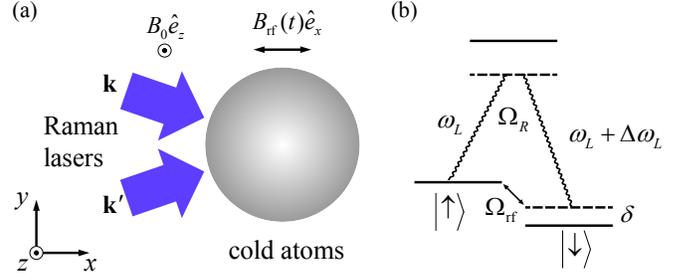}\\
  \caption{(Color online) The experimental setup (a) and energy level diagram (b) used for realizing a 1D Zeeman lattice.
  A uniform bias magnetic field $B_0\hat e_z$ along the $z$ direction fixes the quantization axis. A pair of Raman lasers couple two atomic (pseudo-) spin states with given momentum mismatch.
  A time-dependent rf magnetic field $B_{\text{rf}}(t)\hat{e}_x$ along the $x$ direction couples the two (pseudo-) spin states without momentum exchange. }\label{Fig1}
\end{figure}

The schematic picture of the experimental setup~\cite{Jimenez2012, Cheuk-Zwierlein2012} for creating a 1D Zeeman lattice is shown in Fig.~\ref{Fig1}. A bias magnetic field $B_0{\hat e}_z$ is applied along the $z$ direction to fix the quantization axis. A pair of Raman lasers is applied to synthesize atomic SOC. The (momentum, frequency, polarization) of the two lasers are respectively $(\hbar {\bf{k}}, \omega_L, \pi)$ and $(\hbar {\bf{k}}', \omega_L + \Delta\omega_L, \sigma_+)$, where $\hbar$ is the reduced Planck constant. The two lasers propagate at an intersection angle $\pm \theta_R /2$ with respect to the $x$ axis; they impart momentum $\hbar Q {\hat e}_y = \hbar(\bf{k}' - \bf{k})$ to a single atom accompanied by its up spin flip. Here ${{\hat e}_y}$ is the unit vector along the $y$ direction, and $\hbar Q = 4\pi \hbar \sin(\theta_R /2)/ \lambda_L$ is the magnitude of the two-photon recoil momentum, with $\lambda_L$ the wavelength of the Raman lasers. In Ref.~\cite{Jimenez2012}, the two Raman lasers counter-propagate, i.e., $\theta_R = \pi$, while in Ref.~\cite{Cheuk-Zwierlein2012}, the two lasers propagate at an angle $\theta_R = 38^\circ$. In addition to the pair of Raman lasers, an rf magnetic field along the $x$ direction is applied to drive the transition between atomic spin states without a momentum transfer.

The strength of the bias magnetic field $B_0$ is assumed to be large in the sense that the quadratic Zeeman shift is large enough that only two spin states are effectively coupled by the Raman and rf fields. The single-particle Hamiltonian describing this (pseudo-) spin-$\frac{1}{2}$ system is given by
\begin{equation}
\begin{aligned}
  H = \frac{{{{\mathbf{p}}^2}}}{{2m}} &+ \frac{{\hbar {\omega _Z}}}{2}{\sigma _z} + \hbar {\Omega _{{\text{rf}}}}\cos \left( {{\omega} t + \phi } \right){\sigma _x}  \\
  &  + \hbar {\Omega _R}\cos \left( {\Delta {\omega _L}t + Qy} \right){\sigma _x},  \\
\end{aligned}
\end{equation}
where $\mathbf{p}$ is the atomic momentum, $m$ is the atomic mass,
$\hbar{\omega _Z}$ is the Zeeman splitting of two atomic pseudo-spin states, $\sigma_\mu$ ($\mu = x, y, z$) are the Pauli matrices, ${\Omega}_{{\text{rf}}}$ and ${\Omega _R}$ are the Rabi frequencies for the rf and Raman coupling, respectively,
and ${\omega}$ is the oscillating frequency of the rf field.
Without loss of generality, a relative time phase $\phi$ between the rf and the Raman transition is introduced. Experimentally~\cite{Jimenez2012, Cheuk-Zwierlein2012}, the oscillating frequency of the rf field is
 equal to the frequency difference of Raman lasers, $\omega = \Delta {\omega _L}$.
 In the rotating frame, the Hamiltonian is changed to ${U^\dag }HU - i\hbar {U^\dag }{\partial _t}U$, where $U = \exp \left( { - i{\sigma _z}\omega t/2} \right)$. By omitting the terms oscillating at frequency $2\omega$, i.e., under the rotating wave approximation, we get the time independent Hamiltonian as
\begin{equation}\label{H_y}
  \begin{aligned}
  H = \frac{{{{\mathbf{p}}^2}}}{{2m}} &+  \frac{{\hbar \delta}}{2}{\sigma _z} + \frac{\hbar }{2}\left[ {{\Omega _{{\text{rf}}}}\cos \phi  + {\Omega _R}\cos \left( {Qy} \right)} \right]{\sigma _x}  \\
   & - \frac{\hbar }{2}\left[ {{\Omega _{{\text{rf}}}}\sin \phi  + {\Omega _R}\sin \left( {Qy} \right)} \right]{\sigma _y}, \\
\end{aligned}
\end{equation}
where $\delta = \omega_Z - \omega$ is the detuning from rf (and Raman) resonance.
Equation~(\ref{H_y}) can be cast into the explicit form containing an effective Zeeman field,
\begin{equation}
  H = \frac{{{{\mathbf{p}}^2}}}{{2m}} + {g_F}{\mu _B}{{\mathbf{B}}_{{\text{eff}}}} \cdot \frac{\bm \sigma }{2},
\end{equation}
where $g_F$ is the Land$\acute{\text{e}}$ $g$ factor and $\mu_B$ is the Bohr magneton, and ${{\mathbf{B}}_{{\text{eff}}}} = ( {\Omega _{{\text{rf}}}}\cos \phi  + {\Omega _R}\cos (Qy),\, - {\Omega _{{\text{rf}}}}\sin \phi  - {\Omega _R}\sin (Qy), \,\delta )\times{\hbar }/{{{g_F}{\mu _B}}}$ is the effective Zeeman field, which is spatially periodic along the $y$ direction.
To reveal the implicit SOC in Eq.~(\ref{H_y}), we rewrite the Hamiltonian by carrying out the local space-dependent pseudo-spin rotation operation as has been done in Ref.~\cite{Lin2011} and get
\begin{equation}\label{H_local_SOC}
  \begin{aligned}
  &H= {e^{iQy{\sigma _z}/2}}\{ \frac{{{{\mathbf{p}}^2}}}{{2m}} + \frac{{\hbar Q}}{{2m}}{p_y}{\sigma _z} + \frac{{{E_r}}}{4} + \frac{{\hbar \delta }}{2}{\sigma _z} + \frac{{\hbar {\Omega _R}}}{2}{\sigma _x}  \\
    &\,\,+ \frac{{\hbar {\Omega _{{\text{rf}}}}}}{2}[{\sigma _x}\cos (Qy - \phi ) + {\sigma _y}\sin (Qy - \phi )]\} {e^{ - iQy{\sigma _z}/2}},  \\
\end{aligned}
\end{equation}
where ${E_r} = {\hbar ^2}{Q^2}/2m$ is the two-photon recoil energy.
It is now clear from Eq.~(\ref{H_local_SOC}) that, in the absence of the rf field (by taking ${\Omega _{{\text{rf}}}} = 0$), the Hamiltonian in the rotated frame contains explicit spin ($\sigma_z$) and orbital ($p_y$) coupling~\cite{Lin2011}. When the Raman and rf fields are present at the same time, it is impossible to find a frame transformation to eliminate the lattice term.

In Eq.~(\ref{H_y}), the momentum difference between the two Raman lasers is $\hbar{Q}{{\hat e}_y}$, which leads to the spatially periodic term $\propto\cos(Q y)$. In general, if the direction of the Raman laser pair is rotated in the $x$-$y$ plane to give a momentum difference $\hbar{Q}{{\hat e}_\theta}$ with ${{\hat e}_\theta } = \left( {\cos \theta ,\sin \theta } \right)$, then the spatially periodic term $\cos(Q y)$ in Eq.~(\ref{H_y}) is changed into $\cos(Q {\bf r}\cdot{{\hat e}_\theta })$, where ${\bf r} = (x, y)$ is the 2D atomic coordinate vector.
We further fix the time phase between the rf and Raman transition at $\phi = \pi$; then the corresponding Hamiltonian is given by
\begin{equation}\label{H_theta}
  \begin{aligned}
  H\left( \theta  \right) = \frac{{{{\mathbf{p}}^2}}}{{2m}} &+  \frac{{\hbar \delta}}{2}{\sigma _z} + \frac{{\hbar }}{2}\left[ { - {\Omega _{\text{rf}}}  + {\Omega _R} \cos \left( {Q{\mathbf{r}} \cdot {{\hat e}_\theta }} \right)} \right]{\sigma _x}  \\
   &- \frac{{\hbar {\Omega _R}}}{2}\sin \left( {Q{\mathbf{r}} \cdot {{\hat e}_\theta }} \right){\sigma _y}.  \\
\end{aligned}
\end{equation}
Such a Hamiltonian serves as the building block for realizing topologically nontrivial 2D optical flux lattices, as we explore in the next section.

\section{Dynamical generation of topological optical flux lattices}\label{SecIII}

\begin{figure}[t]
  \includegraphics[width=0.9\columnwidth]{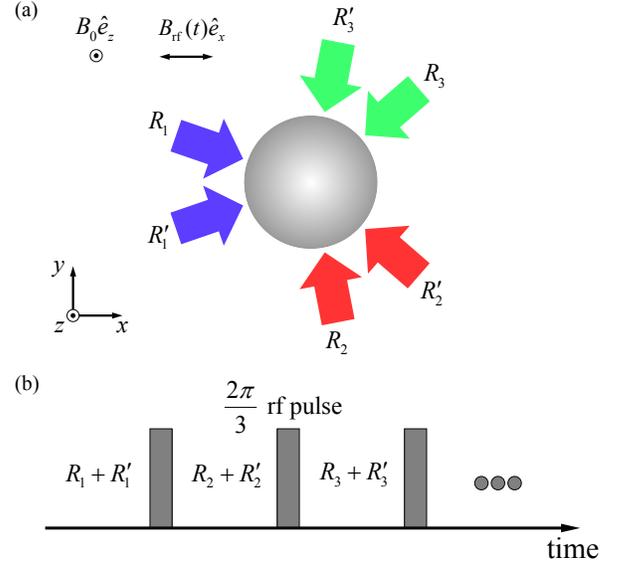}\\
  \caption{(Color online) The setup for realizing an optical flux lattice. (a) The experimental geometry of the bias magnetic field, rf magnetic field, and Raman lasers. (b) The proposed pulse sequence. }\label{Fig2}
\end{figure}

In this section, we extend the scheme on realizing a 1D Zeeman lattice to dynamically generate 2D topological optical flux lattices. The schematic experimental setup is shown in Fig.~\ref{Fig2}, where three pairs of Raman lasers and $2\pi/3$ rf pulses are applied in sequence.
These three pairs of Raman lasers take a mutual angle $120^\circ$, i.e., ${\theta _1} = \frac{\pi }{2}$, ${\theta _2} = \frac{\pi }{2} + \frac{{2\pi }}{3}$ and ${\theta _3} = \frac{\pi }{2} + \frac{{4\pi }}{3}$ [Fig.~\ref{Fig2}(a)]. They are applied to the pancake-shaped cold atoms in three subsequent subperiods, together with three $2\pi/3$ rf pulses which manipulate the spin direction in each evolution period [Fig.~\ref{Fig2}(b)].

The evolution operator for a complete cycle ($T = 3\delta t$) is
\begin{equation}\label{U_OFL}
\begin{aligned}
  U(T,0) =& {e^{i\pi {\sigma _x}/3}}{e^{ - iH\left( {{\theta _3}} \right)\delta t/\hbar }}{e^{i\pi {\sigma _x}/3}}  \\
  &  \times {e^{ - iH\left( {{\theta _2}} \right)\delta t/\hbar }}{e^{i\pi {\sigma _x}/3}}{e^{ - iH\left( {{\theta _1}} \right)\delta t/\hbar }},  \\
\end{aligned}
\end{equation}
where $H\left( {{\theta }} \right)$ is given by Eq.~(\ref{H_theta}).
By repartitioning the rf phase factors, we get
\begin{equation}\label{U_OFL_eff}
  \begin{aligned}
  U(T,0) =& {e^{i\pi {\sigma _x}}} \times \left( {{e^{ - i2\pi {\sigma _x}/3}}{e^{ - iH\left( {{\theta _3}} \right)\delta t/\hbar }}{e^{i2\pi {\sigma _x}/3}}} \right) \hfill \\
  &  \times \left( {{e^{ - i\pi {\sigma _x}/3}}{e^{ - iH\left( {{\theta _2}} \right)\delta t/\hbar }}{e^{i\pi {\sigma _x}/3}}} \right) \times \left( {{e^{ - iH\left( {{\theta _1}} \right)\delta t/\hbar }}} \right)  \\
   =& {e^{ - i{\cal H}\left( {{\theta _3}} \right)\delta t/\hbar }}{e^{ - i{\cal H}\left( {{\theta _2}} \right)\delta t/\hbar }}{e^{ - i{\cal H}\left( {{\theta _1}} \right)\delta t/\hbar }},  \\
\end{aligned}
\end{equation}
where ${{\cal H}}\left( {{\theta }} \right)$ is given by
\begin{equation}\label{H_eff_theta}
\begin{aligned}
   {\cal H}\left({\theta}\right) =& {e^{ - i\left( {\theta  - \pi / 2} \right){\sigma _x}/2}}H\left( \theta  \right){e^{i\left( {\theta  - \pi / 2} \right){\sigma _x}/2}}  \\
   =& \frac{{{{\mathbf{p}}^2}}}{{2m}} + \frac{{\hbar \delta }}{2}\left( {{\sigma _z}\sin \theta  + {\sigma _y}\cos \theta } \right)  \\
   &+ \frac{{\hbar }}{2}\left[ { - {\Omega _{\text{rf}}}  + {\Omega _R} \cos \left( {Q{\mathbf{r}} \cdot {{\hat e}_\theta }} \right)} \right]{\sigma _x}  \\
   & - \frac{{\hbar {\Omega _R}}}{2}\left[\sin \left( {Q{\mathbf{r}} \cdot {{\hat e}_\theta }} \right)\sin \theta\right] {\sigma _y} \\
   &+ \frac{{\hbar {\Omega _R}}}{2}\left[\sin \left( {Q{\mathbf{r}} \cdot {{\hat e}_\theta }} \right)\cos \theta\right] {\sigma _z}.  \\
\end{aligned}
\end{equation}
In the second equal sign of Eq.~(\ref{U_OFL_eff}), a global constant phase factor ${e^{i\pi {\sigma _x}}} =  - 1$  is omitted~\cite{NoteGlobalPhase}.
By defining a time-independent effective Hamiltonian according to $U(T,0) \equiv \exp \left( { - i{H_{{\text{eff}}}}T/\hbar } \right)$, we find to the lowest order of $\delta t$ the effective Hamiltonian
${H_{{\text{eff}}}} \simeq \frac{1}{3}\sum\nolimits_{j = 1}^3 {{\cal H} \left( {{\theta _j}} \right)} $, which can be reformulated as
\begin{equation}\label{H_eff}
  {H_{{\text{eff}}}} = \frac{{{{\mathbf{p}}^2}}}{{2m}} + {\hbar\mathbf{\Omega }} \cdot {\bm{\sigma }}.
\end{equation}
Here, the three components of ${\mathbf{\Omega }}$ are given by [see Fig.~\ref{Fig3}(a) as an illustration]
\begin{equation}\label{Omega_xyz}
  \begin{aligned}
  {\Omega _x} &= \frac{{ {\Omega _R}}}{6}{\left[ - \alpha + \sum\nolimits_{j=1}^{3} {  \cos \left( {Q{\mathbf{r}} \cdot {{\hat e}_j}} \right)} \right]},   \\
  {\Omega _y} &=  - \frac{{ {\Omega _R}}}{6}\sum\nolimits_{j=1}^{3} {\left[ {\sin \left( {Q{\mathbf{r}} \cdot {{\hat e}_j}} \right)\sin {\theta _j}} \right]},   \\
  {\Omega _z} &=   \frac{{ {\Omega _R}}}{6}\sum\nolimits_{j=1}^{3} {\left[ {\sin \left( {Q{\mathbf{r}} \cdot {{\hat e}_j}} \right)\cos {\theta _j}} \right]}, \\
\end{aligned}
\end{equation}
where $\alpha = 3\Omega_{\rm rf}/\Omega_R$, ${\theta _j} = \frac{\pi }{2} + \frac{{2\pi }}{3}\left( {j - 1} \right)$, and ${{\hat e}_j } = \left( {\cos \theta_j ,\sin \theta_j } \right)$. In Eq.~(\ref{H_eff}), the term from finite detuning is absent because $\sum\nolimits_{j = 1}^3 {\cos {\theta _j}}  = 0 = \sum\nolimits_{j = 1}^3 {\sin {\theta _j}} $.

\section{Discussion}\label{SecIV}
The effective Hamiltonian ${H_{{\text{eff}}}}$ realized in the current scheme takes the standard form of an optical flux lattice as proposed by Cooper in Ref.~\cite{Cooper2011}. Generally, an optical flux lattice features large orbital magnetism in the real space, which is often accompanied by topological band structures in the quasimomentum space. The above two points are examined for our effective Hamiltonian Eq.~(\ref{H_eff}) in Secs.~\ref{sub.sec:r} and \ref{sub.sec:k}, respectively. The validity of the effective Hamiltonian approximation is examined for realistic experimental parameters in Sec.~\ref{sub.sec:exp}.

\subsection{General properties of the optical flux lattice}\label{sub.sec:r}

The effective Hamiltonian Eq.~(\ref{H_eff}) contains two terms: the kinetic energy term and the flux lattice term. The flux lattice term resembles the 2D magnetic lattice term of Ref.~\cite{Yu2016}, after a global spin rotation: ${\sigma _x} \to {\sigma _z}$, ${\sigma _y} \to {\sigma _x}$, and ${\sigma _z} \to {\sigma _y}$. Thus similar band topologies are expected to appear at least for the limit that the lattice term dominates the evolution, albeit they have different forms of spatial uniform terms. In Ref.~\cite{Yu2016}, the spatial uniform term takes a spin-orbit coupled form.

The lattice term in Eq.~(\ref{H_eff}) has space-dependent dressed states and effective Zeeman levels,
\begin{equation}
  \hbar {\mathbf{\Omega }} \cdot {\bm\sigma} \left| {{\chi _ \pm ({\mathbf{r}}) }} \right\rangle  = \epsilon_{\pm} ({\mathbf{r}})\left| {{\chi _ \pm ({\mathbf{r}}) }} \right\rangle,
\end{equation}
where $\epsilon_{\pm} ({\mathbf{r}}) = \pm\hbar |{\mathbf{\Omega }}|$.
Generally, the time evolution of the single-particle state is governed by the effective Hamiltonian according to $i\hbar {\partial_t }\left| {\Psi \left( {{\mathbf{r}},{\mathbf{t}}} \right)} \right\rangle  = {H_{{\text{eff}}}}\left| {\Psi \left( {{\mathbf{r}},{\mathbf{t}}} \right)} \right\rangle$, where $\left| {\Psi \left( {{\mathbf{r}},{\mathbf{t}}} \right)} \right\rangle  = \sum\nolimits_{l = +,-} {{\psi _l}\left( {{\mathbf{r}},{\mathbf{t}}} \right)} {\left| {{\chi _l}\left( {\mathbf{r}} \right)} \right\rangle}$.
We define the dimensionless Raman coupling strength as $\beta = \hbar\Omega_R / E_r$.
In the limit of $\hbar\Omega_R \gg E_r$, or $\beta \gg 1$,  the lattice term in Eq.~(\ref{H_eff}) dominates over the kinetic term, thus the low-energy spectrum of $H_{\text{eff}}$ comes from the adiabatic c.m. motion within the lowest dressed state $\left| {{\chi _ {-} } } \right\rangle$ according to~\cite{Dalibard2011}
\begin{equation}\label{Schro_psiMinus}
  i\hbar \frac{\partial }{{\partial t}}{\psi _-} = \left[ {\frac{1}{{2m}}{{\left( {{\mathbf{p}} - {\mathbf{A}}} \right)}^2} + {\epsilon}_- + W} \right]{\psi _-},
\end{equation}
where ${\mathbf{A}} = i\hbar \left\langle {{\chi _ - }|\nabla {\chi _ - }} \right\rangle $ is the geometric vector potential and $W = \frac{{{\hbar ^2}}}{{2m}}|\left\langle {{\chi _ - }|\nabla {\chi _ + }} \right\rangle {|^2}$ is the geometric scalar potential.

\begin{figure}[t]
  \includegraphics[width=0.7\columnwidth]{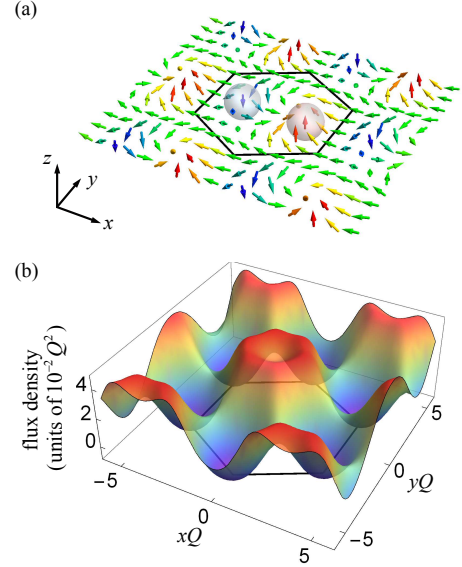}\\
  \caption{(Color online) (a) The three-dimensional view of the local Bloch vector ${\bf m} = {\bf \Omega}/|{\bf \Omega}|$ [with ${\bf \Omega}$ defined in Eq.~(\ref{Omega_xyz})], at $\alpha = 0.9$. The arrows are colored by the magnitude of $m_z$. The hexagon [with edge length $4\pi/(3Q)$] indicates the Wigner-Seitz unit cell of the lattice. Two merons (indicated by two spheres) each with charge $1/2$ appear in the unit cell, which give Skyrmion number $1$. (b) The corresponding flux density $n_\phi$, which gives the net flux over a unit cell $N_\phi = 1$. }\label{Fig3}
\end{figure}

The artificial magnetic flux density ${n_\phi } = \frac{1}{{2\pi \hbar }}{\left( {\nabla  \times {\mathbf{A}}} \right)_z}$ experienced by the optically dressed atoms can be expressed as~\cite{Cooper2011}
\begin{equation}\label{n_phi}
    n_\phi = \frac{1}{4\pi}{{\mathbf{m}} \cdot \left({\partial _x}{\mathbf{m}} \times {\partial _y}{\mathbf{m}} \right)},
\end{equation}
where ${\mathbf{m}} =  - \left\langle {{\chi _ - }} \right|{\bm{\sigma }}\left| {{\chi _ - }} \right\rangle  = {\mathbf{\Omega }}/|{\mathbf{\Omega }}|$ is (minus) the local Bloch vector~\cite{Cooper2011}. Thus, the nontrivial winding pattern of ${\bf m}$ always leads to quantized (nonzero) net flux over the unit cell of the lattice.
The Bloch vector ${\bf m}$ and the flux density $n_\phi$ for our flux lattice are shown in Fig.~\ref{Fig3}. We see from  Fig.~\ref{Fig3}(a) that the local Bloch vector forms a Skyrmion lattice~\cite{Nagaosa2013}, with Skyrmion (winding) number in a unit cell being unity. This Skyrmion number turns out to be equal to the net flux over the unit cell according to Eq.~(\ref{n_phi}), which gives ${N_\phi } = \iint_{{\text{UC}}} {{n_\phi }} = 1$, as is also confirmed by Fig.~\ref{Fig3}(b).

\subsection{Phase diagram}\label{sub.sec:k}

The above description for the artificial gauge fields relies on the adiabatic limit $\beta = \hbar\Omega_R/E_r \gg 1$. Large orbital magnetic field is promising for realizing topological states such as Chern insulators~\cite{Hofstadter1976, Haldane1988}. For general parameter values, the identification of topologically trivial and nontrivial phases deserves a more quantitative description, as we now explore in the following.

\begin{figure}[b]
  \includegraphics[width=0.7\columnwidth]{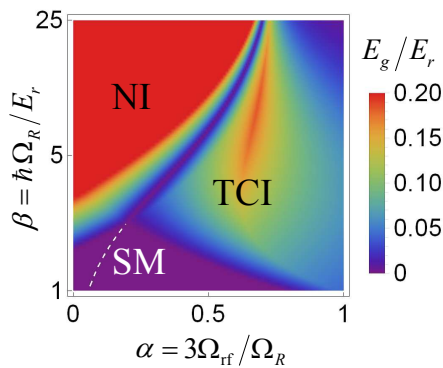}\\
  \caption{(Color online) The lowest energy gap (truncated at $0.2 E_r$) as a function of the Raman and rf field strength. From the band gap and the Chern numbers of the lowest bands, we identify three different phases (assuming unit filling). The two gapped phases, i.e., normal insulator (NI) phase and topological Chern insulator (TCI) phase, are, respectively, with  Chern numbers $C_1 = 0$ and $C_1 = 1$. These two phases are separated by a gapless line ($E_g = 0$), at which the lowest two bands touch. The phase with zero gap is the semimetal (SM) phase. The dashed line in the SM phase indicates a direct band touch for the lowest two bands [$\min \left\{ {{E_2}\left( {\mathbf{q}} \right) - {E_1}\left( {\mathbf{q}} \right)} \right\} = 0$]. Note that a logarithmic (linear) scale is used on the vertical (horizontal) axis. }\label{Fig4}
\end{figure}

To reveal the topological property of the flux lattice system, we need to find the eigenstates and eigenvalues of the effective Hamiltonian $H_{\rm eff}$.
The Bloch wavefunctions ${\psi _{n\mathbf{q}}}\left( {\mathbf{r}} \right)$, labeled with the quasimomentum $q =(q_x, q_y)$ and band index $n$, are the eigenstates of $H_{\rm eff}$, according to the eigenvalue equation
\begin{eqnarray}\label{H_psi_E_psi}
{H_{\rm eff}}{\psi _{n\mathbf{q}}}\left( {\mathbf{r}} \right) = E_n\left( {\mathbf{q}} \right){\psi _{n\mathbf{q}}}\left( {\mathbf{r}} \right),
\end{eqnarray}
where $E_n\left( {\mathbf{q}} \right)$ is the energy spectrum for each quasimomentum restricted to inside the first Brillouin zone [see the inserted
hexagon in Fig.~\ref{Fig6}(a)]. By using the Bloch wavefunctions, we calculate their Berry curvatures $\Omega_n({\mathbf{q}})$ and (first) Chern numbers $C_n$ for each energy band according to~\cite{TKNN1982,Xiao2011}
\begin{eqnarray}
   \begin{gathered}
  {\Omega _n}\left( {\mathbf{q}} \right) = i{\left[ {{\nabla _{\mathbf{q}}} \times \left\langle {{u_n}(\mathbf{q})} \right|{\nabla _{\mathbf{q}}}\left| {{u_n}(\mathbf{q})} \right\rangle } \right]_z}, \hfill \\
  {C_n} = \frac{1}{{2\pi }}\int_{{\text{BZ}}} {{d^2}q{\Omega _n}\left( {\mathbf{q}} \right)} , \hfill \\
\end{gathered}
\end{eqnarray}
where ${u_{n\mathbf{q}}}(\mathbf{r}) = \left\langle {\mathbf{r}\left| {{u_n}(\mathbf{q})} \right.} \right\rangle  = {e^{ - i\mathbf{q} \cdot \mathbf{r}}}{\psi _{n\mathbf{q}}}(\mathbf{r})$ is the cell-periodic part of the Bloch function. A nonzero Chern number indicates quantized transport for a gapped system~\cite{TKNN1982,Dauphin2013,Aidelsburger2015}.

\begin{figure*}[t]
  \includegraphics[width=\textwidth]{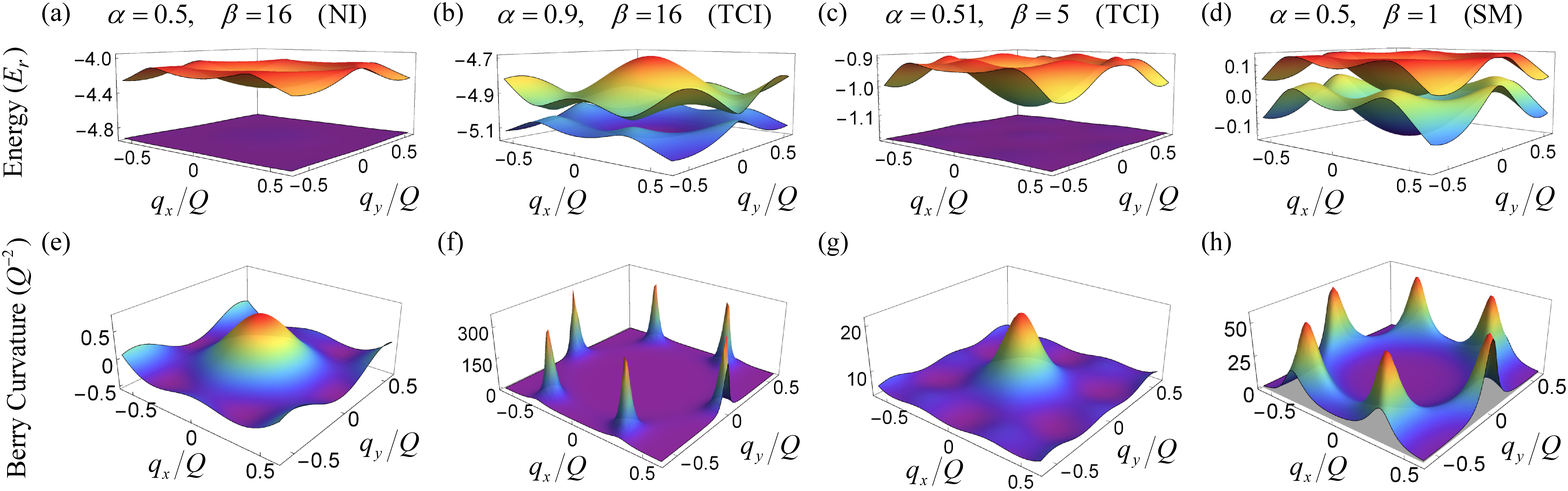}\\
  \caption{(Color online) The energy spectrum for the lowest two bands (a),(b),(c),(d), and the corresponding Berry curvature for the lowest band (e),(f),(g),(h), for the selected $\alpha$ and $\beta$ values.
  The Berry curvature for case (e) gives Chern number $C_1 = 0$, while (f),(g),(h) all correspond to Chern number $C_1 = 1$.}\label{Fig5}
\end{figure*}

We solve Eq.~(\ref{H_psi_E_psi}) using the plane-wave expansion method~\cite{Ashcroft1976} for $H_{\rm eff}$ with different $\alpha$ and $\beta$ values. According to the lowest energy band gap, defined as ${E_g} = \max\left\{  \min \left\{ {{E_2}\left( {\mathbf{q}} \right)} \right\} - \max \left\{ {{E_1}\left( {\mathbf{q}} \right)} \right\} , 0\right\}$ together with the Chern number of the lowest band $C_1$, the ground state phase diagram at unit filling as a function of $\alpha$ and $\beta$ is obtained, which is shown in Fig.~\ref{Fig4}. In the gapped normal insulator (NI) and topological Chern insulator (TCI) phases, this unit filling is equivalent to setting the chemical potential within the lowest energy gap,
leading to a fully filled lowest energy band.
For the gapless semimetal (SM) phase,
the chemical potential lies within the lowest two bands and crosses both of them, leading to both partially filled bands.
The typical band structures of the lowest energy bands and the Berry curvatures for the lowest band for selected ($\alpha$, $\beta$) values are shown in Fig.~\ref{Fig5}.

For the  $\beta\gg1$ regime, i.e., in the adiabatic limit,
two gapped phases with different Chern numbers (i.e., the normal insulator phase with $C_1 = 0$ and the topological Chern insulator phase with $C_1 = 1$) appear.
For a fixed large $\beta$ value, the phase transition from a normal insulator to a topological Chern insulator by varying $\alpha$ can be understood from the tight-binding viewpoint~\cite{Yu2016}. The low-energy spectrum is governed by the Hamiltonian on the right-hand side of Eq.~(\ref{Schro_psiMinus}), which simulates a spinless particle hopping between the lattice sites formed by the minimum of the scalar potential $\epsilon_- + W$ and in the presence of a periodic artificial magnetic field. The scalar potential minima form a simple triangular lattice (with one site per unit cell) at $\alpha = 0$; it changes to a decorated triangular lattice (with three sites per unit cell) near the gapless critical point ($\alpha \sim 0.7$ for $\beta \sim 25$); it then changes to a honeycomb lattice (with two sites per unit cell) for  $\alpha \sim 1$~\footnote{See Fig.~3 in Ref.~\cite{Yu2016} for an illustration of the corresponding scalar potentials.}.
The tight-binding description for the small $\alpha$ case thus involves just one band, and the Chern number for a single tight-binding band is always zero [see Figs.~\ref{Fig5}(a) and~\ref{Fig5}(e) as an example]. While for the $\alpha \sim 1$ case, the honeycomb lattice together with the magnetic flux [shown in Fig.~\ref{Fig3}(b)] realizes the Haldane model~\cite{Haldane1988, Yu2016} of a topological Chern insulator. The energy spectrum $E_{1,2}({\mathbf{q}})$ and Berry curvature $\Omega_1({\mathbf{q}})$ at $\alpha = 0.9$ and $\beta = 16$ are shown  in Figs.~\ref{Fig5}(b) and \ref{Fig5}(f), respectively, from which we can see the opening of an energy
gap at the $\pm K$ points, where the Berry curvature peaks. The lowest two bands for this case take Chern numbers ${C_{1,2}} =  \pm 1$.

In the opposite limit $\beta\sim1$, the kinetic energy term in Eq.~(\ref{H_eff}) dominates over the flux lattice term [the lattice term has a $1/6$ factor as shown in Eq.~(\ref{Omega_xyz})]. In this \emph{weak lattice limit}, the qualitative understanding of the lowest bands can be inferred starting from the free particle spectrum~\cite{Ashcroft1976}. For the spinless case, the spectrum of a particle in the presence of a weak lattice is a folding of the free particle spectrum ${\bf p}^2/2m$, together with a gap opening at the edges of the Brillouin zone due to Bragg reflections~\cite{Ashcroft1976}. For the (quasi-) spin-$\frac{1}{2}$ case as in Eq.~(\ref{H_eff}), the lowest two energy bands in the first Brillouin zone are degenerate in the $\beta\to0$ limit, both taking the free particle spectrum ${\bf q}^2/2m$. A small nonzero $\beta$ breaks this degeneracy, but ${E_2}\left( {{\mathbf{q}} = \Gamma } \right) - {E_1}\left( {{\mathbf{q}} = K} \right) < 0$ is expected for this case as ${E_2}\left( {{\mathbf{q}} = \Gamma } \right)$ is just slightly larger than ${E_1}\left( {{\mathbf{q}} = \Gamma } \right)$. This thus leads to a gapless semimetal state [see Fig.~\ref{Fig5}(d) as an example]. As shown in Fig.~\ref{Fig5}(h) the Berry curvature of the lowest band, the semimetal phase can also take nonzero Chern number $C_1 = 1$. This fact comes from the observation that, by reducing $\beta$ from a gapped topological Chern insulator phase to the semimetal phase, the lowest two bands do not touch; i.e., $\min \left\{ {{E_2}\left( {\mathbf{q}} \right) - {E_1}\left( {\mathbf{q}} \right)} \right\} > 0$ is maintained. Band touching is a necessary condition for changing the Chern numbers of the bands involved; thus, for the region in the right side of the dashed line in the semimetal phase [as shown in Fig.~(\ref{Fig4})], the lowest energy band always takes a Chern number $C_1 = 1$. For a similar reason, $C_1 = 0$ for the left side of the dashed line in the semimetal phase. Although there exists a region with a quantized Chern number $C_1 = 1$ for the semimetal phase, this quantized number does not lead to quantized transport as discussed in Ref.~\cite{TKNN1982}. So we choose not to use the terminologies {\lq\lq}normal semimetal" and {\lq\lq}topological semimetal"~\cite{Goldman2013} to distinguish between these two parameter regions.

For the regime with intermediate $\beta$ (the middle part of the phase diagram), both the adiabatic approximation and the weak lattice argument are inappropriate. Despite 
lacking a simple and proper understanding, topological states with large energy gaps (with a maximal value $\sim0.2E_r$) appear in a large parameter region. Moreover, a nearly nondispersive energy band with Chern number $C_1 = 1$ shows up [see Figs.~\ref{Fig5}(c) and \ref{Fig5}(g) as an example], reminiscent of the lowest Landau level of a charged particle in the presence of a uniform magnetic field~\cite{Cooper2011b}. The lowest band gap is about $20$ times larger than the width of the lowest band for the case shown in Fig.~\ref{Fig5}(c).

\subsection{Experimental parameters and validity of the effective Hamiltonian approximation}\label{sub.sec:exp}

\begin{figure}[htbp]
  \includegraphics[width=\columnwidth]{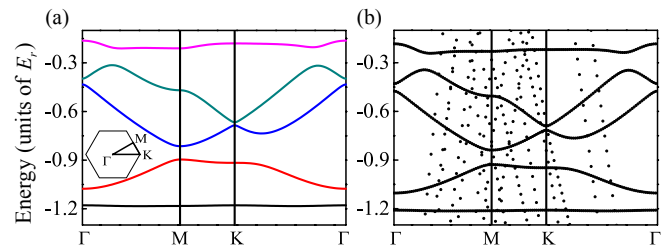}\\
  \caption{(Color online) The energy spectrum along lines with high symmetry in the first Brillouin zone (the inserted hexagon), with $\alpha = 0.51$, $\beta = 5$. (a) Spectrum from the effective Hamiltonian Eq.~(\ref{H_eff}). (b) The quasienergies (shifted by a constant $\hbar \pi/ T$) by evaluating the evolution operator Eq.~(\ref{U_OFL}), with $T  =3\delta t = 15\,\mu \text{s}$.  The value of the two-photon recoil energy is $E_r = (2\pi\hbar)\times1.56\,\rm{kHz}$. }\label{Fig6}
\end{figure}

The scheme we propose to generate an optical flux lattice applies to both bosonic and fermionic atom species. As an especially  interesting result, we find that quasiflat bands with nontrivial topology appear naturally [see Figs.~\ref{Fig5}(c) and \ref{Fig5}(g)], which facilitates fractional fermionic as well as bosonic quantum Hall states when suitable interaction and filling fraction are considered~\cite{Cooper2013}. The experimentally related parameters for realizing an optical flux lattice are as follows: the recoil energy ${E_r} = {\hbar ^2}{Q^2}/2m = 8{\pi ^2}{\hbar ^2}{\sin ^2}({\theta _R}/2)/\left( {m\lambda _L^2} \right)$, the Raman-Rabi frequency ${\Omega _R} = \beta {E_r}/\hbar $, the rf-Rabi frequency ${\Omega _{{\text{rf}}}} = \alpha \beta {E_r}/\left( {3\hbar } \right)$, and the evolution period $T = 3\delta t$.
We consider, for instance, the experimental realization of the topological state with $\alpha = 0.51$, $\beta = 5$ as shown in Figs.~\ref{Fig5}(c) and \ref{Fig5}(g). The recoil energy $E_r$ varies with the atomic mass $m$, the intersection angle between Raman lasers $\theta_R$, and the wavelength of Raman laser $\lambda_L$. For bosonic $^{87}$Rb, with $\theta_R = 38^\circ$ and $\lambda_L = 790\,\text{nm}$, these give rise to $E_r = (2\pi\hbar)\times1.56\,\rm{kHz}$, ${\Omega _R} = (2\pi)\times7.80\,\rm{kHz}$, and ${\Omega _{\rm rf}} = (2\pi)\times1.33\,\rm{kHz}$. To make the effective Hamiltonian Eq.~(\ref{H_eff}) valid for describing time evolution under pulse sequence of Eq.~(\ref{U_OFL}), a small evolution period $T$ is required. By comparing the low-energy spectrum from the effective Hamiltonian Eq.~(\ref{H_eff}) and the Floquet quasienergies from the evolution operator Eq.~(\ref{U_OFL}), as shown in Fig.~\ref{Fig6}, we find that $T = 3\delta t =  15\,\mu \text{s}$ is short enough for validating the effective Hamiltonian description~\footnote{In Fig.~\ref{Fig6}(b), there exist many black dots in addition to the expected band structure similar to the one in Fig.~\ref{Fig6}(a). These dots represent the folding of the quasienergies to the first Floquet-Brillouin zone associated with the high energy part of the static Hamiltonian (with energies exceeding the edges of the first Floquet-Brillouin zone
$\frac{\hbar}{T}\times[-\pi,\pi]$). The states of these black dots are nearly orthogonal to the states that represent the low-frequency physics; thus, the appearance of these dots does not affect the physics of the low-energy effective Hamiltonian.}.
Raman laser pairs can be pulsed on for a duration as short as $\delta t =  5\,\mu \text{s}$,
by using acousto-optic modulators. Thus, all the Raman lasers can be created from a single laser source, and the heating due to spontaneous emission in the current scheme is expected to be comparable to the 1D SOC~\cite{Lin2009, Wang-Zhang2012, Cheuk-Zwierlein2012} or the 1D Zeeman lattices~\cite{Jimenez2012, Cheuk-Zwierlein2012}. By applying the current scheme to the alkaline-earth(-like) atoms~\cite{Mancini_Fallani2015, Song-Jo2016, Livi-Fallani2016} or lanthanide atoms~\cite{Cui-Zhai2013} like Dy~\cite{Nathaniel-Benjamin2016}, such heating can be further suppressed.
Next, we briefly discuss the heating from periodic driving.
The time-averaged effective Hamiltonian in the current scheme is at the zeroth order of the driving period, while the micromotion~\cite{Goldman2014} gives first-order correction to the effective Hamiltonian~\cite{Yu2016}. Thus the heating due to micromotion is suppressed by choosing small driving periods. Adiabatic launching of the lattice strength can further suppress the effects due to micromotion, allowing the ground state of the system to be reached effectively~\cite{Yu2016}.

Before concluding, we would like to point out that, the above consideration does not include the duration of the $2\pi/3$ rf pulse $\delta t'$, which is also of the $\mu \text{s}$ level, and is, in practice, easily accomplished. The evolution during the rf pulse can be expressed as
\begin{equation}
  \exp \left( { - i\frac{{{{\mathbf{p}}^2}}}{{2m\hbar }}\delta t' + i\frac{\pi }{3}{\sigma _x}} \right) = \exp \left( { - i\frac{{{{\mathbf{p}}^2}}}{{2m\hbar }}\delta t'} \right)\exp \left( {i\frac{\pi }{3}{\sigma _x}} \right).
\end{equation}
Thus the finite duration leads to an additional kinetic term in the effective Hamiltonian, which renormalizes $\Omega_R$ in Eq.~(\ref{H_eff}) into $\gamma \Omega_R$, with $\gamma  = \frac{{\delta t}}{{\delta t + \delta t'}}$, while $\alpha$ remains the same.

\section{Conclusion}\label{SecV}

We have presented a scheme to generate topological optical flux lattices using modulated Raman lasers and rf fields. The phase diagram as a function of the Raman and rf field strength is mapped out. Topological quasiflat bands are found to show up naturally in this system, which is promising for studying strongly correlated states and simulating their rich physics within such a simple neutral atom setup. Based on the existing experimental scheme for generating a 1D Zeeman lattice~\cite{Jimenez2012,Cheuk-Zwierlein2012}, the current proposal is highly realizable and the associated intriguing physics is worthy of serious experimental investigations.

\begin{acknowledgments}
This work is supported by the NKBRSFC (Grant No. 2013CB922004) and by the NSFC (Grants No. 91421305, No. 11374176, and No. 11574100). Z.-F. X is supported in part by the National Thousand-Young-Talents Program.
\end{acknowledgments}


\begin{thebibliography}{49}%
\makeatletter
\providecommand \@ifxundefined [1]{%
 \@ifx{#1\undefined}
}%
\providecommand \@ifnum [1]{%
 \ifnum #1\expandafter \@firstoftwo
 \else \expandafter \@secondoftwo
 \fi
}%
\providecommand \@ifx [1]{%
 \ifx #1\expandafter \@firstoftwo
 \else \expandafter \@secondoftwo
 \fi
}%
\providecommand \natexlab [1]{#1}%
\providecommand \enquote  [1]{``#1''}%
\providecommand \bibnamefont  [1]{#1}%
\providecommand \bibfnamefont [1]{#1}%
\providecommand \citenamefont [1]{#1}%
\providecommand \href@noop [0]{\@secondoftwo}%
\providecommand \href [0]{\begingroup \@sanitize@url \@href}%
\providecommand \@href[1]{\@@startlink{#1}\@@href}%
\providecommand \@@href[1]{\endgroup#1\@@endlink}%
\providecommand \@sanitize@url [0]{\catcode `\\12\catcode `\$12\catcode
  `\&12\catcode `\#12\catcode `\^12\catcode `\_12\catcode `\%12\relax}%
\providecommand \@@startlink[1]{}%
\providecommand \@@endlink[0]{}%
\providecommand \url  [0]{\begingroup\@sanitize@url \@url }%
\providecommand \@url [1]{\endgroup\@href {#1}{\urlprefix }}%
\providecommand \urlprefix  [0]{URL }%
\providecommand \Eprint [0]{\href }%
\providecommand \doibase [0]{http://dx.doi.org/}%
\providecommand \selectlanguage [0]{\@gobble}%
\providecommand \bibinfo  [0]{\@secondoftwo}%
\providecommand \bibfield  [0]{\@secondoftwo}%
\providecommand \translation [1]{[#1]}%
\providecommand \BibitemOpen [0]{}%
\providecommand \bibitemStop [0]{}%
\providecommand \bibitemNoStop [0]{.\EOS\space}%
\providecommand \EOS [0]{\spacefactor3000\relax}%
\providecommand \BibitemShut  [1]{\csname bibitem#1\endcsname}%
\let\auto@bib@innerbib\@empty
\bibitem [{\citenamefont {Hasan}\ and\ \citenamefont {Kane}(2010)}]{Hasan2010}%
  \BibitemOpen
  \bibfield  {author} {\bibinfo {author} {\bibfnamefont {M.~Z.}\ \bibnamefont
  {Hasan}}\ and\ \bibinfo {author} {\bibfnamefont {C.~L.}\ \bibnamefont
  {Kane}},\ }\href {\doibase 10.1103/RevModPhys.82.3045} {\bibfield  {journal}
  {\bibinfo  {journal} {Rev. Mod. Phys.}\ }\textbf {\bibinfo {volume} {82}},\
  \bibinfo {pages} {3045} (\bibinfo {year} {2010})}\BibitemShut {NoStop}%
\bibitem [{\citenamefont {Qi}\ and\ \citenamefont {Zhang}(2011)}]{Qi2011}%
  \BibitemOpen
  \bibfield  {author} {\bibinfo {author} {\bibfnamefont {X.-L.}\ \bibnamefont
  {Qi}}\ and\ \bibinfo {author} {\bibfnamefont {S.-C.}\ \bibnamefont {Zhang}},\
  }\href {\doibase 10.1103/RevModPhys.83.1057} {\bibfield  {journal} {\bibinfo
  {journal} {Rev. Mod. Phys.}\ }\textbf {\bibinfo {volume} {83}},\ \bibinfo
  {pages} {1057} (\bibinfo {year} {2011})}\BibitemShut {NoStop}%
\bibitem [{\citenamefont {Madison}\ \emph {et~al.}(2000)\citenamefont
  {Madison}, \citenamefont {Chevy}, \citenamefont {Wohlleben},\ and\
  \citenamefont {Dalibard}}]{Madison2000}%
  \BibitemOpen
  \bibfield  {author} {\bibinfo {author} {\bibfnamefont {K.~W.}\ \bibnamefont
  {Madison}}, \bibinfo {author} {\bibfnamefont {F.}~\bibnamefont {Chevy}},
  \bibinfo {author} {\bibfnamefont {W.}~\bibnamefont {Wohlleben}}, \ and\
  \bibinfo {author} {\bibfnamefont {J.}~\bibnamefont {Dalibard}},\ }\href
  {\doibase 10.1103/PhysRevLett.84.806} {\bibfield  {journal} {\bibinfo
  {journal} {Phys. Rev. Lett.}\ }\textbf {\bibinfo {volume} {84}},\ \bibinfo
  {pages} {806} (\bibinfo {year} {2000})}\BibitemShut {NoStop}%
\bibitem [{\citenamefont {Abo-Shaeer}\ \emph {et~al.}(2001)\citenamefont
  {Abo-Shaeer}, \citenamefont {Raman}, \citenamefont {Vogels},\ and\
  \citenamefont {Ketterle}}]{Abo-Shaeer2001}%
  \BibitemOpen
  \bibfield  {author} {\bibinfo {author} {\bibfnamefont {J.~R.}\ \bibnamefont
  {Abo-Shaeer}}, \bibinfo {author} {\bibfnamefont {C.}~\bibnamefont {Raman}},
  \bibinfo {author} {\bibfnamefont {J.~M.}\ \bibnamefont {Vogels}}, \ and\
  \bibinfo {author} {\bibfnamefont {W.}~\bibnamefont {Ketterle}},\ }\href
  {\doibase 10.1126/science.1060182} {\bibfield  {journal} {\bibinfo  {journal}
  {Science}\ }\textbf {\bibinfo {volume} {292}},\ \bibinfo {pages} {476}
  (\bibinfo {year} {2001})}\BibitemShut {NoStop}%
\bibitem [{\citenamefont {Lin}\ \emph {et~al.}(2009)\citenamefont {Lin},
  \citenamefont {Compton}, \citenamefont {Jimenez-Garcia}, \citenamefont
  {Porto},\ and\ \citenamefont {Spielman}}]{Lin2009}%
  \BibitemOpen
  \bibfield  {author} {\bibinfo {author} {\bibfnamefont {Y.-J.}\ \bibnamefont
  {Lin}}, \bibinfo {author} {\bibfnamefont {R.~L.}\ \bibnamefont {Compton}},
  \bibinfo {author} {\bibfnamefont {K.}~\bibnamefont {Jimenez-Garcia}},
  \bibinfo {author} {\bibfnamefont {J.~V.}\ \bibnamefont {Porto}}, \ and\
  \bibinfo {author} {\bibfnamefont {I.~B.}\ \bibnamefont {Spielman}},\ }\href
  {http://dx.doi.org/10.1038/nature08609} {\bibfield  {journal} {\bibinfo
  {journal} {Nature (London)}\ }\textbf {\bibinfo {volume} {462}},\ \bibinfo
  {pages} {628} (\bibinfo {year} {2009})}\BibitemShut {NoStop}%
\bibitem [{\citenamefont {Lin}\ \emph {et~al.}(2011)\citenamefont {Lin},
  \citenamefont {Jimenez-Garcia},\ and\ \citenamefont {Spielman}}]{Lin2011}%
  \BibitemOpen
  \bibfield  {author} {\bibinfo {author} {\bibfnamefont {Y.-J.}\ \bibnamefont
  {Lin}}, \bibinfo {author} {\bibfnamefont {K.}~\bibnamefont {Jimenez-Garcia}},
  \ and\ \bibinfo {author} {\bibfnamefont {I.~B.}\ \bibnamefont {Spielman}},\
  }\href {http://dx.doi.org/10.1038/nature09887} {\bibfield  {journal}
  {\bibinfo  {journal} {Nature (London)}\ }\textbf {\bibinfo {volume} {471}},\
  \bibinfo {pages} {83} (\bibinfo {year} {2011})}\BibitemShut {NoStop}%
\bibitem [{\citenamefont {Wang}\ \emph {et~al.}(2012)\citenamefont {Wang},
  \citenamefont {Yu}, \citenamefont {Fu}, \citenamefont {Miao}, \citenamefont
  {Huang}, \citenamefont {Chai}, \citenamefont {Zhai},\ and\ \citenamefont
  {Zhang}}]{Wang-Zhang2012}%
  \BibitemOpen
  \bibfield  {author} {\bibinfo {author} {\bibfnamefont {P.}~\bibnamefont
  {Wang}}, \bibinfo {author} {\bibfnamefont {Z.-Q.}\ \bibnamefont {Yu}},
  \bibinfo {author} {\bibfnamefont {Z.}~\bibnamefont {Fu}}, \bibinfo {author}
  {\bibfnamefont {J.}~\bibnamefont {Miao}}, \bibinfo {author} {\bibfnamefont
  {L.}~\bibnamefont {Huang}}, \bibinfo {author} {\bibfnamefont
  {S.}~\bibnamefont {Chai}}, \bibinfo {author} {\bibfnamefont {H.}~\bibnamefont
  {Zhai}}, \ and\ \bibinfo {author} {\bibfnamefont {J.}~\bibnamefont {Zhang}},\
  }\href {\doibase 10.1103/PhysRevLett.109.095301} {\bibfield  {journal}
  {\bibinfo  {journal} {Phys. Rev. Lett.}\ }\textbf {\bibinfo {volume} {109}},\
  \bibinfo {pages} {095301} (\bibinfo {year} {2012})}\BibitemShut {NoStop}%
\bibitem [{\citenamefont {Cheuk}\ \emph {et~al.}(2012)\citenamefont {Cheuk},
  \citenamefont {Sommer}, \citenamefont {Hadzibabic}, \citenamefont {Yefsah},
  \citenamefont {Bakr},\ and\ \citenamefont {Zwierlein}}]{Cheuk-Zwierlein2012}%
  \BibitemOpen
  \bibfield  {author} {\bibinfo {author} {\bibfnamefont {L.~W.}\ \bibnamefont
  {Cheuk}}, \bibinfo {author} {\bibfnamefont {A.~T.}\ \bibnamefont {Sommer}},
  \bibinfo {author} {\bibfnamefont {Z.}~\bibnamefont {Hadzibabic}}, \bibinfo
  {author} {\bibfnamefont {T.}~\bibnamefont {Yefsah}}, \bibinfo {author}
  {\bibfnamefont {W.~S.}\ \bibnamefont {Bakr}}, \ and\ \bibinfo {author}
  {\bibfnamefont {M.~W.}\ \bibnamefont {Zwierlein}},\ }\href {\doibase
  10.1103/PhysRevLett.109.095302} {\bibfield  {journal} {\bibinfo  {journal}
  {Phys. Rev. Lett.}\ }\textbf {\bibinfo {volume} {109}},\ \bibinfo {pages}
  {095302} (\bibinfo {year} {2012})}\BibitemShut {NoStop}%
\bibitem [{\citenamefont {Zhai}(2015)}]{Zhai2015}%
  \BibitemOpen
  \bibfield  {author} {\bibinfo {author} {\bibfnamefont {H.}~\bibnamefont
  {Zhai}},\ }\href {http://stacks.iop.org/0034-4885/78/i=2/a=026001} {\bibfield
   {journal} {\bibinfo  {journal} {Rep. Prog. Phys.}\ }\textbf {\bibinfo
  {volume} {78}},\ \bibinfo {pages} {026001} (\bibinfo {year}
  {2015})}\BibitemShut {NoStop}%
\bibitem [{\citenamefont {Huang}\ \emph {et~al.}(2016)\citenamefont {Huang},
  \citenamefont {Meng}, \citenamefont {Wang}, \citenamefont {Peng},
  \citenamefont {Zhang}, \citenamefont {Chen}, \citenamefont {Li},
  \citenamefont {Zhou},\ and\ \citenamefont {Zhang}}]{Huang-Zhang2016}%
  \BibitemOpen
  \bibfield  {author} {\bibinfo {author} {\bibfnamefont {L.}~\bibnamefont
  {Huang}}, \bibinfo {author} {\bibfnamefont {Z.}~\bibnamefont {Meng}},
  \bibinfo {author} {\bibfnamefont {P.}~\bibnamefont {Wang}}, \bibinfo {author}
  {\bibfnamefont {P.}~\bibnamefont {Peng}}, \bibinfo {author} {\bibfnamefont
  {S.-L.}\ \bibnamefont {Zhang}}, \bibinfo {author} {\bibfnamefont
  {L.}~\bibnamefont {Chen}}, \bibinfo {author} {\bibfnamefont {D.}~\bibnamefont
  {Li}}, \bibinfo {author} {\bibfnamefont {Q.}~\bibnamefont {Zhou}}, \ and\
  \bibinfo {author} {\bibfnamefont {J.}~\bibnamefont {Zhang}},\ }\href
  {http://dx.doi.org/10.1038/nphys3672} {\bibfield  {journal} {\bibinfo
  {journal} {Nat. Phys.}\ }\textbf {\bibinfo {volume} {12}},\ \bibinfo {pages}
  {540} (\bibinfo {year} {2016})}\BibitemShut {NoStop}%
\bibitem [{\citenamefont {Wu}\ \emph {et~al.}(2016)\citenamefont {Wu},
  \citenamefont {Zhang}, \citenamefont {Sun}, \citenamefont {Xu}, \citenamefont
  {Wang}, \citenamefont {Ji}, \citenamefont {Deng}, \citenamefont {Chen},
  \citenamefont {Liu},\ and\ \citenamefont {Pan}}]{USTC_SOC2016}%
  \BibitemOpen
  \bibfield  {author} {\bibinfo {author} {\bibfnamefont {Z.}~\bibnamefont
  {Wu}}, \bibinfo {author} {\bibfnamefont {L.}~\bibnamefont {Zhang}}, \bibinfo
  {author} {\bibfnamefont {W.}~\bibnamefont {Sun}}, \bibinfo {author}
  {\bibfnamefont {X.-T.}\ \bibnamefont {Xu}}, \bibinfo {author} {\bibfnamefont
  {B.-Z.}\ \bibnamefont {Wang}}, \bibinfo {author} {\bibfnamefont {S.-C.}\
  \bibnamefont {Ji}}, \bibinfo {author} {\bibfnamefont {Y.}~\bibnamefont
  {Deng}}, \bibinfo {author} {\bibfnamefont {S.}~\bibnamefont {Chen}}, \bibinfo
  {author} {\bibfnamefont {X.-J.}\ \bibnamefont {Liu}}, \ and\ \bibinfo
  {author} {\bibfnamefont {J.-W.}\ \bibnamefont {Pan}},\ }\href {\doibase
  10.1126/science.aaf6689} {\bibfield  {journal} {\bibinfo  {journal}
  {Science}\ }\textbf {\bibinfo {volume} {354}},\ \bibinfo {pages} {83}
  (\bibinfo {year} {2016})}\BibitemShut {NoStop}%
\bibitem [{\citenamefont {Luo}\ \emph {et~al.}(2016)\citenamefont {Luo},
  \citenamefont {Wu}, \citenamefont {Chen}, \citenamefont {Guan}, \citenamefont
  {Gao}, \citenamefont {Xu}, \citenamefont {You},\ and\ \citenamefont
  {Wang}}]{Luo2015}%
  \BibitemOpen
  \bibfield  {author} {\bibinfo {author} {\bibfnamefont {X.}~\bibnamefont
  {Luo}}, \bibinfo {author} {\bibfnamefont {L.}~\bibnamefont {Wu}}, \bibinfo
  {author} {\bibfnamefont {J.}~\bibnamefont {Chen}}, \bibinfo {author}
  {\bibfnamefont {Q.}~\bibnamefont {Guan}}, \bibinfo {author} {\bibfnamefont
  {K.}~\bibnamefont {Gao}}, \bibinfo {author} {\bibfnamefont {Z.-F.}\
  \bibnamefont {Xu}}, \bibinfo {author} {\bibfnamefont {L.}~\bibnamefont
  {You}}, \ and\ \bibinfo {author} {\bibfnamefont {R.}~\bibnamefont {Wang}},\
  }\href {http://dx.doi.org/10.1038/srep18983} {\bibfield  {journal} {\bibinfo
  {journal} {Sci. Rep.}\ }\textbf {\bibinfo {volume} {6}},\ \bibinfo {pages}
  {18983} (\bibinfo {year} {2016})}\BibitemShut {NoStop}%
\bibitem [{\citenamefont {Struck}\ \emph {et~al.}(2014)\citenamefont {Struck},
  \citenamefont {Simonet},\ and\ \citenamefont
  {Sengstock}}]{Struck_Sengstock2014}%
  \BibitemOpen
  \bibfield  {author} {\bibinfo {author} {\bibfnamefont {J.}~\bibnamefont
  {Struck}}, \bibinfo {author} {\bibfnamefont {J.}~\bibnamefont {Simonet}}, \
  and\ \bibinfo {author} {\bibfnamefont {K.}~\bibnamefont {Sengstock}},\ }\href
  {\doibase 10.1103/PhysRevA.90.031601} {\bibfield  {journal} {\bibinfo
  {journal} {Phys. Rev. A}\ }\textbf {\bibinfo {volume} {90}},\ \bibinfo
  {pages} {031601(R)} (\bibinfo {year} {2014})}\BibitemShut {NoStop}%
\bibitem [{\citenamefont {Jotzu}\ \emph {et~al.}(2015)\citenamefont {Jotzu},
  \citenamefont {Messer}, \citenamefont {G\"org}, \citenamefont {Greif},
  \citenamefont {Desbuquois},\ and\ \citenamefont
  {Esslinger}}]{Jotzu_Esslinger2015}%
  \BibitemOpen
  \bibfield  {author} {\bibinfo {author} {\bibfnamefont {G.}~\bibnamefont
  {Jotzu}}, \bibinfo {author} {\bibfnamefont {M.}~\bibnamefont {Messer}},
  \bibinfo {author} {\bibfnamefont {F.}~\bibnamefont {G\"org}}, \bibinfo
  {author} {\bibfnamefont {D.}~\bibnamefont {Greif}}, \bibinfo {author}
  {\bibfnamefont {R.}~\bibnamefont {Desbuquois}}, \ and\ \bibinfo {author}
  {\bibfnamefont {T.}~\bibnamefont {Esslinger}},\ }\href {\doibase
  10.1103/PhysRevLett.115.073002} {\bibfield  {journal} {\bibinfo  {journal}
  {Phys. Rev. Lett.}\ }\textbf {\bibinfo {volume} {115}},\ \bibinfo {pages}
  {073002} (\bibinfo {year} {2015})}\BibitemShut {NoStop}%
\bibitem [{\citenamefont {Xu}\ \emph {et~al.}(2013)\citenamefont {Xu},
  \citenamefont {You},\ and\ \citenamefont {Ueda}}]{Xu2013}%
  \BibitemOpen
  \bibfield  {author} {\bibinfo {author} {\bibfnamefont {Z.-F.}\ \bibnamefont
  {Xu}}, \bibinfo {author} {\bibfnamefont {L.}~\bibnamefont {You}}, \ and\
  \bibinfo {author} {\bibfnamefont {M.}~\bibnamefont {Ueda}},\ }\href {\doibase
  10.1103/PhysRevA.87.063634} {\bibfield  {journal} {\bibinfo  {journal} {Phys.
  Rev. A}\ }\textbf {\bibinfo {volume} {87}},\ \bibinfo {pages} {063634}
  (\bibinfo {year} {2013})}\BibitemShut {NoStop}%
\bibitem [{\citenamefont {Anderson}\ \emph {et~al.}(2013)\citenamefont
  {Anderson}, \citenamefont {Spielman},\ and\ \citenamefont
  {Juzeli\ifmmode~\bar{u}\else\={u}\fi{}nas}}]{Anderson2013}%
  \BibitemOpen
  \bibfield  {author} {\bibinfo {author} {\bibfnamefont {B.~M.}\ \bibnamefont
  {Anderson}}, \bibinfo {author} {\bibfnamefont {I.~B.}\ \bibnamefont
  {Spielman}}, \ and\ \bibinfo {author} {\bibfnamefont {G.}~\bibnamefont
  {Juzeli\ifmmode~\bar{u}\else\={u}\fi{}nas}},\ }\href {\doibase
  10.1103/PhysRevLett.111.125301} {\bibfield  {journal} {\bibinfo  {journal}
  {Phys. Rev. Lett.}\ }\textbf {\bibinfo {volume} {111}},\ \bibinfo {pages}
  {125301} (\bibinfo {year} {2013})}\BibitemShut {NoStop}%
\bibitem [{\citenamefont {Jimenez-Garcia}\ \emph {et~al.}(2012)\citenamefont
  {Jimenez-Garcia}, \citenamefont {LeBlanc}, \citenamefont {Williams},
  \citenamefont {Beeler}, \citenamefont {Perry},\ and\ \citenamefont
  {Spielman}}]{Jimenez2012}%
  \BibitemOpen
  \bibfield  {author} {\bibinfo {author} {\bibfnamefont {K.}~\bibnamefont
  {Jimenez-Garcia}}, \bibinfo {author} {\bibfnamefont {L.~J.}\ \bibnamefont
  {LeBlanc}}, \bibinfo {author} {\bibfnamefont {R.~A.}\ \bibnamefont
  {Williams}}, \bibinfo {author} {\bibfnamefont {M.~C.}\ \bibnamefont
  {Beeler}}, \bibinfo {author} {\bibfnamefont {A.~R.}\ \bibnamefont {Perry}}, \
  and\ \bibinfo {author} {\bibfnamefont {I.~B.}\ \bibnamefont {Spielman}},\
  }\href {\doibase 10.1103/PhysRevLett.108.225303} {\bibfield  {journal}
  {\bibinfo  {journal} {Phys. Rev. Lett.}\ }\textbf {\bibinfo {volume} {108}},\
  \bibinfo {pages} {225303} (\bibinfo {year} {2012})}\BibitemShut {NoStop}%
\bibitem [{\citenamefont {Cooper}(2011)}]{Cooper2011}%
  \BibitemOpen
  \bibfield  {author} {\bibinfo {author} {\bibfnamefont {N.~R.}\ \bibnamefont
  {Cooper}},\ }\href {\doibase 10.1103/PhysRevLett.106.175301} {\bibfield
  {journal} {\bibinfo  {journal} {Phys. Rev. Lett.}\ }\textbf {\bibinfo
  {volume} {106}},\ \bibinfo {pages} {175301} (\bibinfo {year}
  {2011})}\BibitemShut {NoStop}%
\bibitem [{\citenamefont {Yu}\ \emph {et~al.}(2016)\citenamefont {Yu},
  \citenamefont {Xu}, \citenamefont {L\"u},\ and\ \citenamefont
  {You}}]{Yu2016}%
  \BibitemOpen
  \bibfield  {author} {\bibinfo {author} {\bibfnamefont {J.}~\bibnamefont
  {Yu}}, \bibinfo {author} {\bibfnamefont {Z.-F.}\ \bibnamefont {Xu}}, \bibinfo
  {author} {\bibfnamefont {R.}~\bibnamefont {L\"u}}, \ and\ \bibinfo {author}
  {\bibfnamefont {L.}~\bibnamefont {You}},\ }\href {\doibase
  10.1103/PhysRevLett.116.143003} {\bibfield  {journal} {\bibinfo  {journal}
  {Phys. Rev. Lett.}\ }\textbf {\bibinfo {volume} {116}},\ \bibinfo {pages}
  {143003} (\bibinfo {year} {2016})}\BibitemShut {NoStop}%
\bibitem [{\citenamefont {Hofstadter}(1976)}]{Hofstadter1976}%
  \BibitemOpen
  \bibfield  {author} {\bibinfo {author} {\bibfnamefont {D.~R.}\ \bibnamefont
  {Hofstadter}},\ }\href {\doibase 10.1103/PhysRevB.14.2239} {\bibfield
  {journal} {\bibinfo  {journal} {Phys. Rev. B}\ }\textbf {\bibinfo {volume}
  {14}},\ \bibinfo {pages} {2239} (\bibinfo {year} {1976})}\BibitemShut
  {NoStop}%
\bibitem [{\citenamefont {Dauphin}\ and\ \citenamefont
  {Goldman}(2013)}]{Dauphin2013}%
  \BibitemOpen
  \bibfield  {author} {\bibinfo {author} {\bibfnamefont {A.}~\bibnamefont
  {Dauphin}}\ and\ \bibinfo {author} {\bibfnamefont {N.}~\bibnamefont
  {Goldman}},\ }\href {\doibase 10.1103/PhysRevLett.111.135302} {\bibfield
  {journal} {\bibinfo  {journal} {Phys. Rev. Lett.}\ }\textbf {\bibinfo
  {volume} {111}},\ \bibinfo {pages} {135302} (\bibinfo {year}
  {2013})}\BibitemShut {NoStop}%
\bibitem [{\citenamefont {Aidelsburger}\ \emph {et~al.}(2015)\citenamefont
  {Aidelsburger}, \citenamefont {Lohse}, \citenamefont {Schweizer},
  \citenamefont {Atala}, \citenamefont {Barreiro}, \citenamefont {Nascimbene},
  \citenamefont {Cooper}, \citenamefont {Bloch},\ and\ \citenamefont
  {Goldman}}]{Aidelsburger2015}%
  \BibitemOpen
  \bibfield  {author} {\bibinfo {author} {\bibfnamefont {M.}~\bibnamefont
  {Aidelsburger}}, \bibinfo {author} {\bibfnamefont {M.}~\bibnamefont {Lohse}},
  \bibinfo {author} {\bibfnamefont {C.}~\bibnamefont {Schweizer}}, \bibinfo
  {author} {\bibfnamefont {M.}~\bibnamefont {Atala}}, \bibinfo {author}
  {\bibfnamefont {J.~T.}\ \bibnamefont {Barreiro}}, \bibinfo {author}
  {\bibfnamefont {S.}~\bibnamefont {Nascimbene}}, \bibinfo {author}
  {\bibfnamefont {N.~R.}\ \bibnamefont {Cooper}}, \bibinfo {author}
  {\bibfnamefont {I.}~\bibnamefont {Bloch}}, \ and\ \bibinfo {author}
  {\bibfnamefont {N.}~\bibnamefont {Goldman}},\ }\href
  {http://dx.doi.org/10.1038/nphys3171} {\bibfield  {journal} {\bibinfo
  {journal} {Nat. Phys.}\ }\textbf {\bibinfo {volume} {11}},\ \bibinfo {pages}
  {162} (\bibinfo {year} {2015})}\BibitemShut {NoStop}%
\bibitem [{\citenamefont {Cooper}\ and\ \citenamefont
  {Dalibard}(2013)}]{Cooper2013}%
  \BibitemOpen
  \bibfield  {author} {\bibinfo {author} {\bibfnamefont {N.~R.}\ \bibnamefont
  {Cooper}}\ and\ \bibinfo {author} {\bibfnamefont {J.}~\bibnamefont
  {Dalibard}},\ }\href {\doibase 10.1103/PhysRevLett.110.185301} {\bibfield
  {journal} {\bibinfo  {journal} {Phys. Rev. Lett.}\ }\textbf {\bibinfo
  {volume} {110}},\ \bibinfo {pages} {185301} (\bibinfo {year}
  {2013})}\BibitemShut {NoStop}%
\bibitem [{\citenamefont {Jaksch}\ and\ \citenamefont
  {Zoller}(2003)}]{Jaksch2003}%
  \BibitemOpen
  \bibfield  {author} {\bibinfo {author} {\bibfnamefont {D.}~\bibnamefont
  {Jaksch}}\ and\ \bibinfo {author} {\bibfnamefont {P.}~\bibnamefont
  {Zoller}},\ }\href {\doibase 10.1088/1367-2630/5/1/356} {\bibfield  {journal}
  {\bibinfo  {journal} {New J. Phys.}\ }\textbf {\bibinfo {volume} {5}},\
  \bibinfo {pages} {56} (\bibinfo {year} {2003})}\BibitemShut {NoStop}%
\bibitem [{\citenamefont {Celi}\ \emph {et~al.}(2014)\citenamefont {Celi},
  \citenamefont {Massignan}, \citenamefont {Ruseckas}, \citenamefont {Goldman},
  \citenamefont {Spielman}, \citenamefont {Juzeli\ifmmode~\bar{u}\else
  \={u}\fi{}nas},\ and\ \citenamefont {Lewenstein}}]{Celi-Lewenstein2014}%
  \BibitemOpen
  \bibfield  {author} {\bibinfo {author} {\bibfnamefont {A.}~\bibnamefont
  {Celi}}, \bibinfo {author} {\bibfnamefont {P.}~\bibnamefont {Massignan}},
  \bibinfo {author} {\bibfnamefont {J.}~\bibnamefont {Ruseckas}}, \bibinfo
  {author} {\bibfnamefont {N.}~\bibnamefont {Goldman}}, \bibinfo {author}
  {\bibfnamefont {I.~B.}\ \bibnamefont {Spielman}}, \bibinfo {author}
  {\bibfnamefont {G.}~\bibnamefont {Juzeli\ifmmode~\bar{u}\else
  \={u}\fi{}nas}}, \ and\ \bibinfo {author} {\bibfnamefont {M.}~\bibnamefont
  {Lewenstein}},\ }\href {\doibase 10.1103/PhysRevLett.112.043001} {\bibfield
  {journal} {\bibinfo  {journal} {Phys. Rev. Lett.}\ }\textbf {\bibinfo
  {volume} {112}},\ \bibinfo {pages} {043001} (\bibinfo {year}
  {2014})}\BibitemShut {NoStop}%
\bibitem [{\citenamefont {Aidelsburger}\ \emph {et~al.}(2013)\citenamefont
  {Aidelsburger}, \citenamefont {Atala}, \citenamefont {Lohse}, \citenamefont
  {Barreiro}, \citenamefont {Paredes},\ and\ \citenamefont
  {Bloch}}]{Aidelsburger2013}%
  \BibitemOpen
  \bibfield  {author} {\bibinfo {author} {\bibfnamefont {M.}~\bibnamefont
  {Aidelsburger}}, \bibinfo {author} {\bibfnamefont {M.}~\bibnamefont {Atala}},
  \bibinfo {author} {\bibfnamefont {M.}~\bibnamefont {Lohse}}, \bibinfo
  {author} {\bibfnamefont {J.~T.}\ \bibnamefont {Barreiro}}, \bibinfo {author}
  {\bibfnamefont {B.}~\bibnamefont {Paredes}}, \ and\ \bibinfo {author}
  {\bibfnamefont {I.}~\bibnamefont {Bloch}},\ }\href {\doibase
  10.1103/PhysRevLett.111.185301} {\bibfield  {journal} {\bibinfo  {journal}
  {Phys. Rev. Lett.}\ }\textbf {\bibinfo {volume} {111}},\ \bibinfo {pages}
  {185301} (\bibinfo {year} {2013})}\BibitemShut {NoStop}%
\bibitem [{\citenamefont {Miyake}\ \emph {et~al.}(2013)\citenamefont {Miyake},
  \citenamefont {Siviloglou}, \citenamefont {Kennedy}, \citenamefont {Burton},\
  and\ \citenamefont {Ketterle}}]{Miyake2013}%
  \BibitemOpen
  \bibfield  {author} {\bibinfo {author} {\bibfnamefont {H.}~\bibnamefont
  {Miyake}}, \bibinfo {author} {\bibfnamefont {G.~A.}\ \bibnamefont
  {Siviloglou}}, \bibinfo {author} {\bibfnamefont {C.~J.}\ \bibnamefont
  {Kennedy}}, \bibinfo {author} {\bibfnamefont {W.~C.}\ \bibnamefont {Burton}},
  \ and\ \bibinfo {author} {\bibfnamefont {W.}~\bibnamefont {Ketterle}},\
  }\href {\doibase 10.1103/PhysRevLett.111.185302} {\bibfield  {journal}
  {\bibinfo  {journal} {Phys. Rev. Lett.}\ }\textbf {\bibinfo {volume} {111}},\
  \bibinfo {pages} {185302} (\bibinfo {year} {2013})}\BibitemShut {NoStop}%
\bibitem [{\citenamefont {Kennedy}\ \emph {et~al.}(2015)\citenamefont
  {Kennedy}, \citenamefont {Burton}, \citenamefont {Chung},\ and\ \citenamefont
  {Ketterle}}]{Kennedy_Ketterle2015}%
  \BibitemOpen
  \bibfield  {author} {\bibinfo {author} {\bibfnamefont {C.~J.}\ \bibnamefont
  {Kennedy}}, \bibinfo {author} {\bibfnamefont {W.~C.}\ \bibnamefont {Burton}},
  \bibinfo {author} {\bibfnamefont {W.~C.}\ \bibnamefont {Chung}}, \ and\
  \bibinfo {author} {\bibfnamefont {W.}~\bibnamefont {Ketterle}},\ }\href
  {\doibase 10.1038/nphys3421} {\bibfield  {journal} {\bibinfo  {journal} {Nat.
  Phys.}\ }\textbf {\bibinfo {volume} {11}},\ \bibinfo {pages} {859} (\bibinfo
  {year} {2015})}\BibitemShut {NoStop}%
\bibitem [{\citenamefont {Mancini}\ \emph {et~al.}(2015)\citenamefont
  {Mancini}, \citenamefont {Pagano}, \citenamefont {Cappellini}, \citenamefont
  {Livi}, \citenamefont {Rider}, \citenamefont {Catani}, \citenamefont {Sias},
  \citenamefont {Zoller}, \citenamefont {Inguscio}, \citenamefont {Dalmonte},\
  and\ \citenamefont {Fallani}}]{Mancini_Fallani2015}%
  \BibitemOpen
  \bibfield  {author} {\bibinfo {author} {\bibfnamefont {M.}~\bibnamefont
  {Mancini}}, \bibinfo {author} {\bibfnamefont {G.}~\bibnamefont {Pagano}},
  \bibinfo {author} {\bibfnamefont {G.}~\bibnamefont {Cappellini}}, \bibinfo
  {author} {\bibfnamefont {L.}~\bibnamefont {Livi}}, \bibinfo {author}
  {\bibfnamefont {M.}~\bibnamefont {Rider}}, \bibinfo {author} {\bibfnamefont
  {J.}~\bibnamefont {Catani}}, \bibinfo {author} {\bibfnamefont
  {C.}~\bibnamefont {Sias}}, \bibinfo {author} {\bibfnamefont {P.}~\bibnamefont
  {Zoller}}, \bibinfo {author} {\bibfnamefont {M.}~\bibnamefont {Inguscio}},
  \bibinfo {author} {\bibfnamefont {M.}~\bibnamefont {Dalmonte}}, \ and\
  \bibinfo {author} {\bibfnamefont {L.}~\bibnamefont {Fallani}},\ }\href
  {\doibase 10.1126/science.aaa8736} {\bibfield  {journal} {\bibinfo  {journal}
  {Science}\ }\textbf {\bibinfo {volume} {349}},\ \bibinfo {pages} {1510}
  (\bibinfo {year} {2015})}\BibitemShut {NoStop}%
\bibitem [{\citenamefont {Stuhl}\ \emph {et~al.}(2015)\citenamefont {Stuhl},
  \citenamefont {Lu}, \citenamefont {Aycock}, \citenamefont {Genkina},\ and\
  \citenamefont {Spielman}}]{Stuhl_Spielman2015}%
  \BibitemOpen
  \bibfield  {author} {\bibinfo {author} {\bibfnamefont {B.~K.}\ \bibnamefont
  {Stuhl}}, \bibinfo {author} {\bibfnamefont {H.-I.}\ \bibnamefont {Lu}},
  \bibinfo {author} {\bibfnamefont {L.~M.}\ \bibnamefont {Aycock}}, \bibinfo
  {author} {\bibfnamefont {D.}~\bibnamefont {Genkina}}, \ and\ \bibinfo
  {author} {\bibfnamefont {I.~B.}\ \bibnamefont {Spielman}},\ }\href {\doibase
  10.1126/science.aaa8515} {\bibfield  {journal} {\bibinfo  {journal}
  {Science}\ }\textbf {\bibinfo {volume} {349}},\ \bibinfo {pages} {1514}
  (\bibinfo {year} {2015})}\BibitemShut {NoStop}%
\bibitem [{Note1()}]{Note1}%
  \BibitemOpen
  \bibinfo {note} {The terminology \protect \emph {magnetic lattice} is
  basically equivalent to \protect \emph {Zeeman lattice}. Some may use the
  former to emphasize that such a lattice is generated
  magnetically.}\BibitemShut {Stop}%
\bibitem [{\citenamefont {Luo}\ \emph {et~al.}(2015)\citenamefont {Luo},
  \citenamefont {Wu}, \citenamefont {Chen}, \citenamefont {Lu}, \citenamefont
  {Wang},\ and\ \citenamefont {You}}]{Luo2014}%
  \BibitemOpen
  \bibfield  {author} {\bibinfo {author} {\bibfnamefont {X.}~\bibnamefont
  {Luo}}, \bibinfo {author} {\bibfnamefont {L.}~\bibnamefont {Wu}}, \bibinfo
  {author} {\bibfnamefont {J.}~\bibnamefont {Chen}}, \bibinfo {author}
  {\bibfnamefont {R.}~\bibnamefont {Lu}}, \bibinfo {author} {\bibfnamefont
  {R.}~\bibnamefont {Wang}}, \ and\ \bibinfo {author} {\bibfnamefont
  {L.}~\bibnamefont {You}},\ }\href
  {http://stacks.iop.org/1367-2630/17/i=8/a=083048} {\bibfield  {journal}
  {\bibinfo  {journal} {New J. Phys.}\ }\textbf {\bibinfo {volume} {17}},\
  \bibinfo {pages} {083048} (\bibinfo {year} {2015})}\BibitemShut {NoStop}%
\bibitem [{\citenamefont {Cooper}\ and\ \citenamefont
  {Dalibard}(2011)}]{Cooper2011b}%
  \BibitemOpen
  \bibfield  {author} {\bibinfo {author} {\bibfnamefont {N.~R.}\ \bibnamefont
  {Cooper}}\ and\ \bibinfo {author} {\bibfnamefont {J.}~\bibnamefont
  {Dalibard}},\ }\href {\doibase 10.1209/0295-5075/95/66004} {\bibfield
  {journal} {\bibinfo  {journal} {Europhys. Lett.}\ }\textbf {\bibinfo {volume}
  {95}},\ \bibinfo {pages} {66004} (\bibinfo {year} {2011})}\BibitemShut
  {NoStop}%
\bibitem [{\citenamefont {Juzeli\ifmmode~\bar{u}\else \={u}\fi{}nas}\ and\
  \citenamefont {Spielman}(2012)}]{Juzeliunas2012}%
  \BibitemOpen
  \bibfield  {author} {\bibinfo {author} {\bibfnamefont {G.}~\bibnamefont
  {Juzeli\ifmmode~\bar{u}\else \={u}\fi{}nas}}\ and\ \bibinfo {author}
  {\bibfnamefont {I.}~\bibnamefont {Spielman}},\ }\href {\doibase
  10.1088/1367-2630/14/12/123022} {\bibfield  {journal} {\bibinfo  {journal}
  {New J. Phys.}\ }\textbf {\bibinfo {volume} {14}},\ \bibinfo {pages} {123022}
  (\bibinfo {year} {2012})}\BibitemShut {NoStop}%
\bibitem [{\citenamefont {Thouless}\ \emph {et~al.}(1982)\citenamefont
  {Thouless}, \citenamefont {Kohmoto}, \citenamefont {Nightingale},\ and\
  \citenamefont {den Nijs}}]{TKNN1982}%
  \BibitemOpen
  \bibfield  {author} {\bibinfo {author} {\bibfnamefont {D.~J.}\ \bibnamefont
  {Thouless}}, \bibinfo {author} {\bibfnamefont {M.}~\bibnamefont {Kohmoto}},
  \bibinfo {author} {\bibfnamefont {M.~P.}\ \bibnamefont {Nightingale}}, \ and\
  \bibinfo {author} {\bibfnamefont {M.}~\bibnamefont {den Nijs}},\ }\href
  {\doibase 10.1103/PhysRevLett.49.405} {\bibfield  {journal} {\bibinfo
  {journal} {Phys. Rev. Lett.}\ }\textbf {\bibinfo {volume} {49}},\ \bibinfo
  {pages} {405} (\bibinfo {year} {1982})}\BibitemShut {NoStop}%
\bibitem [{Not()}]{NoteGlobalPhase}%
  \BibitemOpen
  \href@noop {} {}\bibinfo {note} {This global uniform phase factor just leads
  to a constant energy shift $\hbar\pi/T$ in the effective Hamiltonian
  description.}\BibitemShut {Stop}%
\bibitem [{\citenamefont {Dalibard}\ \emph {et~al.}(2011)\citenamefont
  {Dalibard}, \citenamefont {Gerbier}, \citenamefont
  {Juzeli\ifmmode~\bar{u}\else \={u}\fi{}nas},\ and\ \citenamefont
  {\"Ohberg}}]{Dalibard2011}%
  \BibitemOpen
  \bibfield  {author} {\bibinfo {author} {\bibfnamefont {J.}~\bibnamefont
  {Dalibard}}, \bibinfo {author} {\bibfnamefont {F.}~\bibnamefont {Gerbier}},
  \bibinfo {author} {\bibfnamefont {G.}~\bibnamefont
  {Juzeli\ifmmode~\bar{u}\else \={u}\fi{}nas}}, \ and\ \bibinfo {author}
  {\bibfnamefont {P.}~\bibnamefont {\"Ohberg}},\ }\href {\doibase
  10.1103/RevModPhys.83.1523} {\bibfield  {journal} {\bibinfo  {journal} {Rev.
  Mod. Phys.}\ }\textbf {\bibinfo {volume} {83}},\ \bibinfo {pages} {1523}
  (\bibinfo {year} {2011})}\BibitemShut {NoStop}%
\bibitem [{\citenamefont {Nagaosa}\ and\ \citenamefont
  {Tokura}(2013)}]{Nagaosa2013}%
  \BibitemOpen
  \bibfield  {author} {\bibinfo {author} {\bibfnamefont {N.}~\bibnamefont
  {Nagaosa}}\ and\ \bibinfo {author} {\bibfnamefont {Y.}~\bibnamefont
  {Tokura}},\ }\href {http://dx.doi.org/10.1038/nnano.2013.243} {\bibfield
  {journal} {\bibinfo  {journal} {Nat. Nanotechnol.}\ }\textbf {\bibinfo
  {volume} {8}},\ \bibinfo {pages} {899} (\bibinfo {year} {2013})}\BibitemShut
  {NoStop}%
\bibitem [{\citenamefont {Haldane}(1988)}]{Haldane1988}%
  \BibitemOpen
  \bibfield  {author} {\bibinfo {author} {\bibfnamefont {F.~D.~M.}\
  \bibnamefont {Haldane}},\ }\href {\doibase 10.1103/PhysRevLett.61.2015}
  {\bibfield  {journal} {\bibinfo  {journal} {Phys. Rev. Lett.}\ }\textbf
  {\bibinfo {volume} {61}},\ \bibinfo {pages} {2015} (\bibinfo {year}
  {1988})}\BibitemShut {NoStop}%
\bibitem [{\citenamefont {Xiao}\ \emph {et~al.}(2010)\citenamefont {Xiao},
  \citenamefont {Chang},\ and\ \citenamefont {Niu}}]{Xiao2011}%
  \BibitemOpen
  \bibfield  {author} {\bibinfo {author} {\bibfnamefont {D.}~\bibnamefont
  {Xiao}}, \bibinfo {author} {\bibfnamefont {M.-C.}\ \bibnamefont {Chang}}, \
  and\ \bibinfo {author} {\bibfnamefont {Q.}~\bibnamefont {Niu}},\ }\href
  {\doibase 10.1103/RevModPhys.82.1959} {\bibfield  {journal} {\bibinfo
  {journal} {Rev. Mod. Phys.}\ }\textbf {\bibinfo {volume} {82}},\ \bibinfo
  {pages} {1959} (\bibinfo {year} {2010})}\BibitemShut {NoStop}%
\bibitem [{\citenamefont {Ashcroft}\ and\ \citenamefont
  {Mermin}(1976)}]{Ashcroft1976}%
  \BibitemOpen
  \bibfield  {author} {\bibinfo {author} {\bibfnamefont {N.~W.}\ \bibnamefont
  {Ashcroft}}\ and\ \bibinfo {author} {\bibfnamefont {N.~D.}\ \bibnamefont
  {Mermin}},\ }\href@noop {} {\emph {\bibinfo {title} {Solid State Physics}}}\
  (\bibinfo  {publisher} {Cengage Learning},\ \bibinfo {address} {New York},\
  \bibinfo {year} {1976})\BibitemShut {NoStop}%
\bibitem [{Note2()}]{Note2}%
  \BibitemOpen
  \bibinfo {note} {See Fig.~3 in Ref.~\cite {Yu2016} for an illustration of the
  corresponding scalar potentials.}\BibitemShut {Stop}%
\bibitem [{\citenamefont {Goldman}\ \emph {et~al.}(2013)\citenamefont
  {Goldman}, \citenamefont {Anisimovas}, \citenamefont {Gerbier}, \citenamefont
  {\"Ohberg}, \citenamefont {Spielman},\ and\ \citenamefont
  {Juzeli\ifmmode~\bar{u}\else\={u}\fi{}nas}}]{Goldman2013}%
  \BibitemOpen
  \bibfield  {author} {\bibinfo {author} {\bibfnamefont {N.}~\bibnamefont
  {Goldman}}, \bibinfo {author} {\bibfnamefont {E.}~\bibnamefont {Anisimovas}},
  \bibinfo {author} {\bibfnamefont {F.}~\bibnamefont {Gerbier}}, \bibinfo
  {author} {\bibfnamefont {P.}~\bibnamefont {\"Ohberg}}, \bibinfo {author}
  {\bibfnamefont {I.~B.}\ \bibnamefont {Spielman}}, \ and\ \bibinfo {author}
  {\bibfnamefont {G.}~\bibnamefont
  {Juzeli\ifmmode~\bar{u}\else\={u}\fi{}nas}},\ }\href {\doibase
  10.1088/1367-2630/15/1/013025} {\bibfield  {journal} {\bibinfo  {journal}
  {New J. Phys.}\ }\textbf {\bibinfo {volume} {15}},\ \bibinfo {pages} {013025}
  (\bibinfo {year} {2013})}\BibitemShut {NoStop}%
\bibitem [{Note3()}]{Note3}%
  \BibitemOpen
  \bibinfo {note} {In Fig.~\ref {Fig6}(b), there exist many black dots in
  addition to the expected band structure similar to the one in Fig.~\ref
  {Fig6}(a). These dots represent the folding of the quasienergies to the first
  Floquet-Brillouin zone associated with the high energy part of the static
  Hamiltonian (with energies exceeding the edges of the first Floquet-Brillouin
  zone $\protect \frac {\hbar }{T}\times [-\pi ,\pi ]$). The states of these
  black dots are nearly orthogonal to the states that represent the
  low-frequency physics; thus, the appearance of these dots does not affect the
  physics of the low-energy effective Hamiltonian.}\BibitemShut {Stop}%
\bibitem [{\citenamefont {Song}\ \emph {et~al.}(2016)\citenamefont {Song},
  \citenamefont {He}, \citenamefont {Zhang}, \citenamefont {Hajiyev},
  \citenamefont {Huang}, \citenamefont {Liu},\ and\ \citenamefont
  {Jo}}]{Song-Jo2016}%
  \BibitemOpen
  \bibfield  {author} {\bibinfo {author} {\bibfnamefont {B.}~\bibnamefont
  {Song}}, \bibinfo {author} {\bibfnamefont {C.}~\bibnamefont {He}}, \bibinfo
  {author} {\bibfnamefont {S.}~\bibnamefont {Zhang}}, \bibinfo {author}
  {\bibfnamefont {E.}~\bibnamefont {Hajiyev}}, \bibinfo {author} {\bibfnamefont
  {W.}~\bibnamefont {Huang}}, \bibinfo {author} {\bibfnamefont {X.-J.}\
  \bibnamefont {Liu}}, \ and\ \bibinfo {author} {\bibfnamefont {G.-B.}\
  \bibnamefont {Jo}},\ }\href {\doibase 10.1103/PhysRevA.94.061604} {\bibfield
  {journal} {\bibinfo  {journal} {Phys. Rev. A}\ }\textbf {\bibinfo {volume}
  {94}},\ \bibinfo {pages} {061604(R)} (\bibinfo {year} {2016})}\BibitemShut
  {NoStop}%
\bibitem [{\citenamefont {Livi}\ \emph {et~al.}(2016)\citenamefont {Livi},
  \citenamefont {Cappellini}, \citenamefont {Diem}, \citenamefont {Franchi},
  \citenamefont {Clivati}, \citenamefont {Frittelli}, \citenamefont {Levi},
  \citenamefont {Calonico}, \citenamefont {Catani}, \citenamefont {Inguscio},\
  and\ \citenamefont {Fallani}}]{Livi-Fallani2016}%
  \BibitemOpen
  \bibfield  {author} {\bibinfo {author} {\bibfnamefont {L.~F.}\ \bibnamefont
  {Livi}}, \bibinfo {author} {\bibfnamefont {G.}~\bibnamefont {Cappellini}},
  \bibinfo {author} {\bibfnamefont {M.}~\bibnamefont {Diem}}, \bibinfo {author}
  {\bibfnamefont {L.}~\bibnamefont {Franchi}}, \bibinfo {author} {\bibfnamefont
  {C.}~\bibnamefont {Clivati}}, \bibinfo {author} {\bibfnamefont
  {M.}~\bibnamefont {Frittelli}}, \bibinfo {author} {\bibfnamefont
  {F.}~\bibnamefont {Levi}}, \bibinfo {author} {\bibfnamefont {D.}~\bibnamefont
  {Calonico}}, \bibinfo {author} {\bibfnamefont {J.}~\bibnamefont {Catani}},
  \bibinfo {author} {\bibfnamefont {M.}~\bibnamefont {Inguscio}}, \ and\
  \bibinfo {author} {\bibfnamefont {L.}~\bibnamefont {Fallani}},\ }\href
  {\doibase 10.1103/PhysRevLett.117.220401} {\bibfield  {journal} {\bibinfo
  {journal} {Phys. Rev. Lett.}\ }\textbf {\bibinfo {volume} {117}},\ \bibinfo
  {pages} {220401} (\bibinfo {year} {2016})}\BibitemShut {NoStop}%
\bibitem [{\citenamefont {Cui}\ \emph {et~al.}(2013)\citenamefont {Cui},
  \citenamefont {Lian}, \citenamefont {Ho}, \citenamefont {Lev},\ and\
  \citenamefont {Zhai}}]{Cui-Zhai2013}%
  \BibitemOpen
  \bibfield  {author} {\bibinfo {author} {\bibfnamefont {X.}~\bibnamefont
  {Cui}}, \bibinfo {author} {\bibfnamefont {B.}~\bibnamefont {Lian}}, \bibinfo
  {author} {\bibfnamefont {T.-L.}\ \bibnamefont {Ho}}, \bibinfo {author}
  {\bibfnamefont {B.~L.}\ \bibnamefont {Lev}}, \ and\ \bibinfo {author}
  {\bibfnamefont {H.}~\bibnamefont {Zhai}},\ }\href {\doibase
  10.1103/PhysRevA.88.011601} {\bibfield  {journal} {\bibinfo  {journal} {Phys.
  Rev. A}\ }\textbf {\bibinfo {volume} {88}},\ \bibinfo {pages} {011601(R)}
  (\bibinfo {year} {2013})}\BibitemShut {NoStop}%
\bibitem [{\citenamefont {Burdick}\ \emph {et~al.}(2016)\citenamefont
  {Burdick}, \citenamefont {Tang},\ and\ \citenamefont
  {Lev}}]{Nathaniel-Benjamin2016}%
  \BibitemOpen
  \bibfield  {author} {\bibinfo {author} {\bibfnamefont {N.~Q.}\ \bibnamefont
  {Burdick}}, \bibinfo {author} {\bibfnamefont {Y.}~\bibnamefont {Tang}}, \
  and\ \bibinfo {author} {\bibfnamefont {B.~L.}\ \bibnamefont {Lev}},\ }\href
  {\doibase 10.1103/PhysRevX.6.031022} {\bibfield  {journal} {\bibinfo
  {journal} {Phys. Rev. X}\ }\textbf {\bibinfo {volume} {6}},\ \bibinfo {pages}
  {031022} (\bibinfo {year} {2016})}\BibitemShut {NoStop}%
\bibitem [{\citenamefont {Goldman}\ and\ \citenamefont
  {Dalibard}(2014)}]{Goldman2014}%
  \BibitemOpen
  \bibfield  {author} {\bibinfo {author} {\bibfnamefont {N.}~\bibnamefont
  {Goldman}}\ and\ \bibinfo {author} {\bibfnamefont {J.}~\bibnamefont
  {Dalibard}},\ }\href {\doibase 10.1103/PhysRevX.4.031027} {\bibfield
  {journal} {\bibinfo  {journal} {Phys. Rev. X}\ }\textbf {\bibinfo {volume}
  {4}},\ \bibinfo {pages} {031027} (\bibinfo {year} {2014})}\BibitemShut
  {NoStop}%
\end{thebibliography}
%

\end{document}